\documentclass[11pt,a4paper]{article}
\usepackage{jheppub}
\usepackage{enumitem} 
\usepackage{lmodern}
\usepackage[utf8]{inputenc}
\usepackage{color}
\usepackage[dvipsnames, svgnames, x11names]{xcolor}
\usepackage{float}
\usepackage{amsmath,bm}
\usepackage{amssymb}
\usepackage{graphicx}
\usepackage{setspace}
\usepackage{subfigure}
\usepackage{amsthm}
\usepackage{amsfonts}
\usepackage{braket}
\usepackage{multirow}
\usepackage{ulem}
\usepackage{slashed}
\usepackage{babel}
\usepackage{lipsum}
\usepackage{hyperref}
\usepackage{url}
\usepackage{footnote}
\usepackage{indentfirst}
\usepackage{booktabs}
\usepackage{array}

\hypersetup{
	colorlinks=true,
	linkcolor=cyan,
	urlcolor=blue,
	citecolor=red,
}

\def\p{\partial}

\def\r{\rho}
\def\({\left(}
\def\){\right)}
\def\L{\left}
\def\R{\right}
\abstract{In this paper, we investigate the microscopic derivation of the entropy and the holographic RG flow in 3D C-metric. We first discuss the case of a sector in BTZ (Banados-Teitelboim-Zanelli) black hole. By rescaling the Newton's constant we recover the area law of entropy of this sector by microstate counting. Then we apply this technique to all accelerating BTZ phases in 3D C-metric. Finally, for the boundary entropy in 3D C-metric, we study the monotonicity of the $g$-function of 3D C-metric in small acceleration limit and find that  the $g$-theorem is satisfied  only in $\rm I_{2}$. }

\title{Thermodynamics and Holographic RG Flow in 3D C-metric}
\author[a]{Shaohua Xue}
\author[b]{, Yingyu Yang}
\author[a]{and Li-Xin Li}
\affiliation[a]{The Kavli Institute for Astronomy and Astrophysics, Peking University,
	Beijing 100871, China}
\affiliation[b]{Center for High Energy Physics, Peking University, No.5 Yiheyuan Rd, Beijing 100871, P. R.
	China}
\emailAdd{xuesh@pku.edu.cn, yangyingyu@pku.edu.cn, lxl@pku.edu.cn}

\begin{document}
	\maketitle
	
	\section{Introduction}
	After many aspects of investigations in four-dimensional C-metric \cite{levi1918ds2,weyl1919statischen,kinnersley1970uniformly,bonnor1983sources,griffiths2006interpreting,plebanski1976rotating,PhysRevD.2.1359,dias2003pair,emparan2000exact2,emparan2000exact}, a natural promotion is to discuss the C-metric in three dimensions by truncating one dimension  \cite{arenas2022acceleration,Arenas-Henriquez:2023hur,Cisterna:2023qhh} to explore the quantum and holographic features more conveniently. In AdS$_{3}$, it takes a form as:
	\begin{equation}\label{3dcmetric}
		\mathrm{d}s^2=\frac{1}{A^2(x-y)^2}\left[-P(y)\mathrm{d}\tau^2+\frac{1}{P(y)}\mathrm{d}y^2+ \frac{1}{Q(x)}\mathrm{d}x^2\right],
	\end{equation}
	with negative cosmological constant $-\frac{\ 1}{\ l^2}$ and acceleration $A$. According to the Lorentzian signature, \eqref{3dcmetric} has three distinct solutions named Class I, Class II, Class III:
	\begin{equation}
		\begin{cases}\label{C-metric}
			\text{Class I : } & P(y)=\frac{1}{A^2 l^2}+y^2-1, \  Q(x)=1-x^2,\  \text{with}\  |x|\leq1,\\
			\\
			\text{Class II : } & P(y)=\frac{1}{A^2 l^2}+1-y^2, \  Q(x)=x^2-1,\  \text{with}\  |x|\geq1,\\
			\\
			\text{Class III : } & P(y)=\frac{1}{A^2 l^2}-(1+y^2), \  Q(x)=x^2+1,\  \text{with}\  x\in\mathbb{R}.
		\end{cases}
	\end{equation}
	As we can see, the $y$ direction is an analogue of the radial direction, whereas the $x$ direction can not  always be directly compared to the angular direction, since in some cases the conformal boundary is partly masked by the horizon of the black hole phase, which will cause divergences in mass and entropy. Therefore, there is a must to insert end-of-the-world (EOW) branes at different $x$ to truncate the range of $x$, and the sectors cut from these three solutions have two different mass interpretations, i.e. an accelerating particle in thermal AdS generated by a conical defect and an accelerating BTZ black hole generated by the quotient of the spacetime. The tension of the brane at $x=X$ is given by Israel equations \cite{israel1966singular}
	\begin{align}
		4\pi \sigma h_{ij}&=4\pi\int^{+}_{-}T_{ij}=K_{ij}-Kh_{ij},\\
		\sigma&=\pm\frac{\ A}{\ 4\pi}\sqrt{Q(X)},
	\end{align}
	where $K_{ij}$ is the extrinsic curvature of the $x=X$ surface and the $h_{ij}$ is the corresponding induced metric. The plus (minus) sign corresponds to the domain (strut) wall \cite{arenas2022acceleration}. The  EOW branes constrain  the boundary field theory, and generate defects coupled to it  \cite{Arenas-Henriquez:2023hur,tian2024aspects}. Thus the analysis of C-metric can be divided into two parts: defect geometry and bounded theory.

	The geometry of defects has been studied in \cite{fursaev1995description} and can be obtained by metric rescaling \cite{de2011near}, where the field theory does not couple to the defect. The investigation of the reconstruction reveals that defects contain hidden degrees of freedom and  can be recovered by a lifting map\cite{balasubramanian2015entwinement}. In AdS/CFT correspondence, the defect has attracted increasing attention from both sides \cite{,,benjamin2020pure,kastikainen2020conical,belin2022spectrum,smolkin2015correlation,arefeva2015holographic}, and it can be interpreted as heavy operators inserted in the boundary field theory. Recently, we notice that in \cite{grabovsky2024heavy}, the promotion of the angular defect bridges the gap between particles and  BTZ black holes. Specifically, above the critical threshold mass $M_{\rm vac}=-\frac{\ 1}{\ 8G_{3}}$ for vacuum AdS, a conical defect can be viewed as a black hole with an imaginary temperature, and conversely  a BTZ black hole can be thought of as a conical geometry with an imaginary angular deficit. That is to say, the conical AdS (particle) and the black hole can both be regarded as defect theories sourced by different heavy operators on the boundary. 
	
	On the other hand, the development of AdS/BCFT provides a holographic frame to investigate the defect which is coupled to the boundary \cite{Cardy:2004hm}. One of the starting points is to focus on the holographic RG flow and the holographic $g$-theorem. In \cite{Affleck:1991tk}, the $g$-function can be defined from the boundary entropy and monotonically decreases under the boundary RG flow, which is proven in\cite{Friedan:2003yc}. Then the holographic RG is investigated in 5D $\mathcal{N}=8$ super-gravity \cite{Freedman:1999gp}, after which the holographic $g$-theorem get a test in D-branes \cite{Yamaguchi:2002pa}. Based on the previous results, $g$-theorem is further employed in Randall-Sundrum braneworld gravity and used to discuss the BCFT\cite{Takayanagi:2011zk}, after which the $g$-theorem gets investigated in AdS/BCFT \cite{Casini:2016fgb,Flory:2017ftd}.
	
	In 3D C-metric, all phases have topological defects in the geometry and the corresponding boundary field theories are coupled to them. Accordingly, we can investigate these phases from two perspectives. In the first perspective, we can analyze the geometric defect and investigate the thermodynamics by counting the microstates with the asymptotic algebra. According to the AdS/CFT correspondence, all the information can be  encoded on the boundary, and  the microscopic derivation of the entropy of BTZ black holes in the large mass limit supports this conjecture \cite{strominger1998black}. Thus we manage to give an algebraic description of part of the entropy of C-metric. In the second perspective, we  discuss the boundary entropy and investigate the holographic RG flow in weak coupling with small acceleration to gain further insights on the boundary field theory.
	
	This paper is organized as follows: In Section \ref{btzbh}, we  first review AdS$_{3}$ geometry in Chern-Simons formalism and the computation of the microscopic states of the BTZ black hole. Then we discuss how to count the microstates of a sector in BTZ black hole, and recover the area law of entropy. Next in Section \ref{acbhc}, we apply the method to three-dimensional C-metric and  reproduce the area law of entropy, then we consider the holographic RG flow in small acceleration limit and find the corresponding $g$-function in all three solutions. Finally, we draw our conclusions in Section \ref{conclusion}.

	\section{BTZ black hole} \label{btzbh}
	It is well known that the three-dimensional AdS gravity can be reformulated as a  Chern-Simons (CS) gauge theory with the action:
	\begin{equation}\label{2.1}
		I_{\rm CS}(A_{\pm})=\frac{\ k}{\ 4\pi} \int \left\langle A_{+}\wedge \mathrm{d}A_{+}+\frac{\ 2}{\ 3}A_{+}\wedge A_{+}\wedge A_{+} \right\rangle-\frac{\ k}{\ 4\pi} \int \left\langle A_{-}\wedge \mathrm{d}A_{-}+\frac{\ 2}{\ 3}A_{-}\wedge A_{-}\wedge A_{-} \right\rangle,
	\end{equation}
	where $A_{\pm}$ is the gauge connection in $sl(2,\mathbb{R})$ and $k$ is the CS level
	\begin{align}
		k=\frac{\ c}{\ 6}.
	\end{align}
	Plus and minus correspond to two chiral modes. The geometry of the asymptotic $\rm AdS_{3}$ can be obtained from the Banados geometry with the cosmological constant $\Lambda=-\frac{\ 1}{\ l^2}$ \cite{banados1999three,sheikh20163d}:
	\begin{equation}
		\begin{aligned}\label{Banandos geometries}
			\mathrm{d}s^2&=\mathrm{d}\rho^2-\left(e^{2\rho/l}+l^{4}e^{-2\rho/l}\mathcal{L}_{+}(x^{+})\mathcal{L}_{-}(x^{-})\right)\mathrm{d}x^{+}\mathrm{d}x^{-}+l^2\mathcal{L}_{-}(x^{-})(\mathrm{d}x^{-})^2+l^2\mathcal{L}_{+}(x^{+})(\mathrm{d}x^{+})^2,\\
			x^{\pm}&=t\pm l\phi,        
		\end{aligned}
	\end{equation}
	where $\mathcal{L}_{+}(x^{+})$ and $\mathcal{L}_{-}(x^{-})$ are two arbitrary functions. In CS theory, the metric \eqref{Banandos geometries} can be expressed by gauge connections as
	\begin{equation}
		g_{\mu\nu}=\frac{l^{2}}{2}\left\langle \left(A_{\mu(+)}-A_{\mu(-)}\right)\left(A_{\nu(+)}-A_{\nu(-)}\right)\right\rangle ,
	\end{equation}
	with the same trace in \eqref{2.1} and 
	\begin{equation}
		A_{\pm}=e^{\mp \frac{\r}{l}T_{0}}\left[\mathrm{d}\pm\left(T_{\pm 1}-\mathcal{L}_{\pm}(x^{\pm})T_{\mp 1}\right)\frac{\ \mathrm{d}x^{\pm}}{\ l} \right]e^{\pm \frac{\r}{l}T_{0}},
	\end{equation}
	where $T_{0}$ and $T_{\pm 1}$ are the generators of $sl(2,\mathbb{R})$. 
	Given a precise boundary condition, the CS theory has two towers of conical boundary charge:
	\begin{align}
		\delta Q(\epsilon^{+})&=\frac{\ k}{\ 2\pi}\oint \left\langle \epsilon^{+}\delta A_{+}\right\rangle,\label{Charge1} \\
		\delta Q(\epsilon^{-})&=\frac{\ k}{\ 2\pi}\oint\left\langle \epsilon^{-}\delta A_{-}\right\rangle,\label{Charge2}
	\end{align}
	where $\delta A$ is the gauge connection fluctuation allowed by the boundary condition and is equal to the deformation caused by the gauge transformation generated by $\epsilon^{\pm}$:
	\begin{equation}\label{T1}
		\delta_{\epsilon^{\pm}}A_{\pm}=\mathcal{O}(\delta A_{\pm}).
	\end{equation}
	The parameter $\epsilon^{\pm}$ comprises all the boundary conditions preserving \eqref{T1} and it can be  parameterized by two factors $\lambda^{\pm}(x^{\pm})$ as:
	\begin{equation}
		\epsilon^{\pm}=e^{\pm \frac{\r}{l}L_{0}}\left[ \lambda^{\pm}(x^{\pm})L_{\pm 1}-l\lambda^{\pm}(x^{\pm})'L_{0}+\left(\frac{\ l^2}{\ 2}\lambda^{\pm}(x^{\pm})''-\mathcal{L}_{\pm}\left(x^{\pm}\lambda^{\pm}(x^{\pm})\right)\right)L_{\mp 1}\right]e^{\mp \frac{\r}{l}L_{0}}.\label{boundcond}
	\end{equation}
	Equivalently, $\lambda^{\pm}(x^{\pm})$ parameterizes all allowed asymptotic symmetries and boundary charges.
	Then the variation of charges \eqref{Charge1} and \eqref{Charge2} gives the surface charges
	\begin{align}
		Q^{+}&=\frac{\ k l}{\ 2\pi}\oint \lambda^{+}\mathcal{L}_{+}(x^{+})\mathrm{d}\phi,\label{Charge3}\\
		Q^{-}&=\frac{\ k l}{\ 2\pi}\oint \lambda^{-}\mathcal{L}_{-}(x^{-})\mathrm{d}\phi.\label{Charge4}
	\end{align}
	Given $\lambda_{1}^{\pm}(x^{\pm})$ and $\lambda_{2}^{\pm}(x^{\pm})$, the Poisson bracket of surface charges gives:
	\begin{equation}
		\begin{aligned}
			i\left\{Q^{\pm}(\lambda^{\pm}_{1}(x^{\pm}), Q^{\pm}(\lambda^{\pm}_{2}(x^{\pm})\right \}=&Q^{\pm}(\lambda^{\pm}_{1}(x^{\pm})\lambda^{\pm}_{2}(x^{\pm})'-\lambda^{\pm}_{1}(x^{\pm})\lambda^{\pm}_{2}(x^{\pm})')\\
			&+\frac{\ kl}{\ 4\pi}\oint \lambda^{\pm}_{1}(x^{\pm})'''\lambda^{\pm}_{2}(x^{\pm})\mathrm{d}\phi .
		\end{aligned}
	\end{equation}
	We can make the expressions of $Q^{\pm}$ as Fourier transforms by letting $\lambda^{\pm}_{n}(x^{\pm})=le^{-inx^{\pm}/l}$ ($n\in\mathbb{Z}$), then \eqref{Charge3} and \eqref{Charge4} gives:
	\begin{align}\label{2.5}
		L_{n}^{+}=&Q^{+}(e^{-inx^{+}/l})=\frac{\ kl}{\ 2\pi}\oint \lambda^{+}_{n}\mathcal{L}_{+}(x^{+})\mathrm{d}\phi ,\\
		L_{n}^{-}=&Q^{+}(e^{-inx^{-}/l})=\frac{\ kl}{\ 2\pi}\oint \lambda^{-}_{n}\mathcal{L}_{-}(x^{-})\mathrm{d}\phi .
	\end{align}
	From the commutation relations of the surface charges we can find  $L^{\pm}_{n}$ forms the Virasoro algebra after a shift $L_{0}\rightarrow L_{0}-\frac{\ c}{\ 24}$:
	\begin{align}
		\left[ L_{n}^{+}, L_{m}^{+}  \right]&=(n-m)L_{n+m}^{+}+\frac{\ c(n^3-n)}{\ 12}\delta_{n+m,0},\label{V1}\\
		\left[ L_{n}^{-}, L_{m}^{-}  \right]&=(n-m)L_{n+m}^{-}+\frac{\ c(n^3-n)}{\ 12}\delta_{n+m,0},\label{V2}\\
		\left[ L_{n}^{+}, L_{m}^{-}  \right]&=0,\label{V3}
	\end{align}
	where the shift $E_{\rm vac}=-\frac{\ c}{\ 12}$ is the Casimir energy for the ground state.
	
	In the following we first review the BTZ black hole and then discuss the sector cut from the complete geometry of it.

	\subsection{A Complete BTZ black hole}
	The line element of a complete BTZ black hole is:
	\begin{equation}\label{BTZ}
		\mathrm{d}s^2=-\left(\frac{\ r^2}{\ l^2}-8G_{3}m\right)\mathrm{d}t^2+\frac{\ 1}{\ \frac{\ r^2}{\ l^2}-8G_{3}m}\mathrm{d}r^2+r^2\mathrm{d}\phi^2,
	\end{equation}
	with the period $\phi\sim \phi+2\pi$ and $l$ the 3D radius. The temperature can be written as 
	\begin{align}
		T_{\rm h}=\frac{\ 1}{\ \beta}=\frac{\ \sqrt{2G_{3}m}}{\ \pi l}.
	\end{align}
	To facilitate our use of the conclusion in the Chern-Simons theory, we rewrite the line element by this transformation: 
	\begin{align}
		r=\sqrt{8G_{3}m}\, l\cosh(\frac{\ \rho}{\ l}), \quad t=\frac{\ x^{+}+x^{-}}{\ 2\sqrt{2G_{3}m}}, \quad \phi=\frac{\ x^{+}-x^{-}}{\ 2\sqrt{2G_{3}m}\,l},
	\end{align}
	then \eqref{BTZ} becomes:
	\begin{equation}\label{BTZmetric}
		\mathrm{d}s^2=\mathrm{d}\rho^2-\left(e^{\frac{\ 2\r}{\ l}}+e^{-\frac{\ 2\r}{\ l}}\right)\mathrm{d}x^{+}\mathrm{d}x^{-}+(\mathrm{d}x^{+})^2+(\mathrm{d}x^{-})^2.
	\end{equation}
	Automatically, we can read off the constants $\mathcal{L}_{+}(x^{+})$ and $\mathcal{L}_{-}(x^{-})$ as $\frac{ 1}{l^2}$ by comparing \eqref{Banandos geometries} and \eqref{BTZmetric}. Accordingly, the two light-cone parameters are 
	\begin{align}
		x^{\pm}=\sqrt{2G_{3}m}\,t\pm l\sqrt{2G_{3}m}\,\phi,
	\end{align}
	with the period 
	\begin{align}
		x^{\pm}\sim x^{\pm}+2\pi l\sqrt{2G_{3}m}.
	\end{align}
	Therefore the  quantities  which parametrize all allowed asymptotic symmetries in \eqref{boundcond} can be Fourier expanded by: 
	\begin{align}
		\lambda^{+}_{n}=\sqrt{2G_{3}m}\,l\, e^{\frac{\ -inx^{+}}{\ \sqrt{2G_{3}m}\,l}}, \quad \lambda^{-}_{n}=\sqrt{2G_{3}m}\,l \, e^{\frac{\ -inx^{-}}{\ \sqrt{2G_{3}m}\,l}}.
	\end{align}
	The conical surface charges $Q^{\pm}$ generated by $\lambda^{\pm}_{n}$  construct the Virasoro algebra in \eqref{V1}, \eqref{V2} and \eqref{V3}. Moreover,
	the zero mode gives the mass:
	\begin{align}\label{0 mode2}
		L_{0}^{+}=\frac{\ k}{\ 2\pi}\oint \lambda^{+}_{0}\mathcal{L}_{+}(x^{+})\sqrt{2G_{3}m}\,l\mathrm{d}\phi&=\frac{\ ml}{\ 2}\\
		L_{0}^{-}=\frac{\ k}{\ 2\pi}\oint \lambda^{-}_{0}\mathcal{L}_{-}(x^{-})\sqrt{2G_{3}m}\,l\mathrm{d}\phi&=\frac{\ ml}{\ 2}.\\
		M_{\rm BTZ}=\frac{ L_{0}^{+}+L_{0}^{-}}{\ l}&=m,\\
		L_{0}^{+}-L_{0}^{-}&=0.
	\end{align}
	Given the large mass ($ml\gg c$), by using the Cardy formula which is used in CFT to count the microstate \cite{cardy1986operator}, the entropy of the BTZ balck hole is recovered as shown in \cite{strominger1998black}:
	\begin{equation}
		\begin{aligned}\label{BH2}
			S_{\rm BTZ}&=2\pi\sqrt{\frac{\ c}{\ 6}L_{0}^{+}}+2\pi\sqrt{\frac{\ c}{\ 6}L_{0}^{-}},\\
			&=\frac{\ A}{\ 4G_{3}},\\
			&=S_{\rm BH}.
		\end{aligned}
	\end{equation}
	The result is in agreement with Bekenstein-Hawking (BH) entropy. 
	Here for the sake of clarity we have made a distinction, that $S_{\rm BH}$ is the Bekenstein-Hawking entropy while $S_{\rm BTZ}$ is the result of the Cardy formula with $G_{3}$ and $m$.

	\subsection{A Sector of BTZ Black Hole}
\begin{figure}[h]
		\centering
		\includegraphics[width=1.0\linewidth]{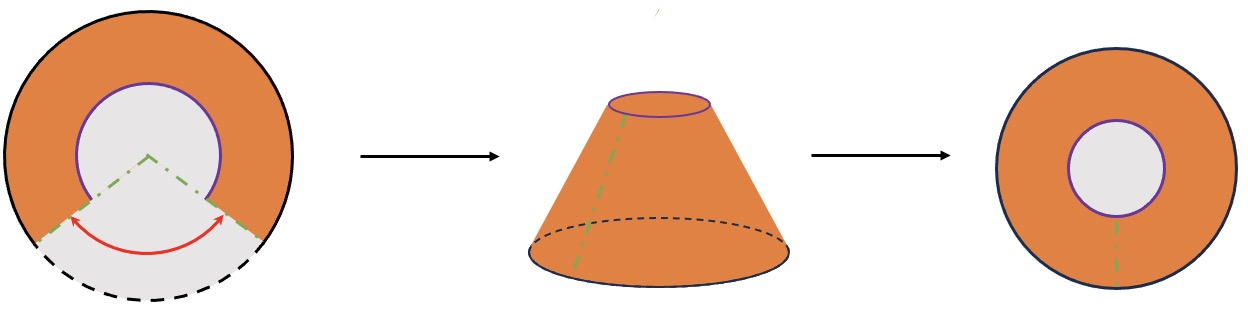}
		\caption{The sector (orange region) wrapped by branes (green dash-dot lines), horizon (purple curves) and asymptotic boundary (black curves). \textbf{Left}: The sector is part of the complete interior, and the grey region is removed. \textbf{Middle}: The surface of a truncated cone is constructed by gluing the sector along the branes and identifying $\phi$ with $\phi+2\pi \alpha$. \textbf{Right}: After rescaling the angular direction and recovering the angular range from $0$ to $2\pi$, the sector becomes an annulus.}
		\label{CAdS}
	\end{figure}

	In this part we consider a sector that is cut out from the complete geometry of a BTZ black hole. In global coordinates, the line element of this sector is represented as \cite{Banados:1992wn}:
	\begin{equation}\label{3.1}
		\mathrm{d}s^2=-\left(\frac{\ r^2}{\ l^2}-8G_{3}m\right)\mathrm{d}t^2+\frac{\ 1}{\ \frac{\ r^2}{\ l^2}-8G_{3}m}\mathrm{d}r^2+r^2\mathrm{d}\phi^2,
	\end{equation}
	with $\phi$ ranging from $\phi_{0}$ to $\phi_{0}+2\pi \alpha$, and $\alpha$ denotes the deficit angular and $\alpha\in (0,1]$. To facilitate our use of the conclusion in Chern-Simons theory, we rewrite the line element by the same transformation as in the previous section.  But the integration of the conical surface charges is taken in $(0,2\pi \alpha)$, only part of the circle, as is illustrated in Fig. \ref{CAdS} (Left). Then \eqref{0 mode2} gives:
	\begin{align}
		\frac{\ k}{\ 2\pi}\int \lambda^{+}_{0}\mathcal{L}_{+}(x^{+})\sqrt{2G_{3}m}\,l\mathrm{d}\phi=\alpha L_{0}^{+}&=\frac{\ \alpha ml}{\ 2}, \label{0 mode3}\\
		\frac{\ k}{\ 2\pi}\int \lambda^{-}_{0}\mathcal{L}_{-}(x^{-})\sqrt{2G_{3}m}\,l\mathrm{d}\phi=\alpha L_{0}^{-}&=\frac{\ \alpha ml}{\ 2},\\
		M_{\rm cBTZ}=\frac{ L_{0}^{+}+L_{0}^{-}}{\ l}&=\alpha m=\alpha M_{\rm BTZ},\\
		L_{0}^{+}-L_{0}^{-}&=0. \label{0 mode32}
	\end{align} 
	From \eqref{0 mode3}-\eqref{0 mode32}, $\alpha M_{\rm BTZ}$ corresponds to the mass of the sector cut out from the entire geometry of BTZ black hole. However, there is a problem with this process: The parameter $\lambda$ in the boundary condition is expanded over the period of $\phi$, $( 0, 2 \pi)$, but the integral of the conical surface charge is not extended over a full period, and thus this approach cannot define a Viraoso algebra. To solve this problem we can recover the period $( 0, 2 \pi)$ in the angular direction by a transformation that glues the two branes by identifying $\phi_{0}$ with $\phi_{0}+2\pi\alpha$ , as shown in Fig. \ref{CAdS} (Middle). 
	Following \cite{de2011near}, by the following transformation 
	\begin{align}
		\phi=\alpha \tilde{\phi}, \quad r=\frac{\ \tilde{r}}{\ \alpha}, \quad t=\alpha\tilde{t},
	\end{align}
	we can rewrite the line element of \eqref{3.1}:
	\begin{equation}\label{CBTZ}
		\mathrm{d}s^2=-\left(\frac{\ \tilde{r}^2}{\ l^2}-8\tilde{G_{3}}\tilde{m}\right)\mathrm{d}\tilde{t}^2+\frac{\ 1}{\ \frac{\ \tilde{r}^2}{\ l^2}-8\tilde{G_{3}}\tilde{m}}\mathrm{d}\tilde{r}^2+\tilde{r}^2\mathrm{d}\tilde{\phi}^2,  
	\end{equation}
	with $\tilde{G}_{3}=\alpha G_{3}$, $\tilde{m}=\alpha m$ and $\tilde{\phi}\sim\tilde{\phi}+2\pi$. This is to say, we can use a new gravity theory to describe the conical AdS. And in this new theory, the angular direction is restored to the range from 0 to $2\pi$, as  shown in Fig. $\ref{CAdS}$ (Right). To make it clear, in the following discussion, the phrase `old theory' refers to the theory with Newton's constant $G_{3}$ and mass $m$, and the phrase `new theory' refers to the theory with Newton's constant $\tilde{G}_{3}$ and mass $\tilde{m}$. In the new gravity theory, the temperature is 
	\begin{align}
		\tilde{T}_{h}=\frac{\ 1}{\ \tilde{\beta}}=\frac{\ \sqrt{2\tilde{G}_{3}\tilde{m}}}{\ \pi l}=\frac{\ \alpha}{\ \beta}.
	\end{align}
	Following the above procedure, accordingly we have:
	\begin{align}
		\tilde{r}=\sqrt{8\tilde{G_{3}}\tilde{m}}\, l\cosh(\frac{ \rho}{ l}), \quad \tilde{t}=\frac{\ x^{+}+x^{-}}{\ 2\sqrt{2\tilde{G_{3}}\tilde{m}}}, \quad \tilde{\phi}=\frac{\ x^{+}-x^{-}}{\ 2\sqrt{2\tilde{G_{3}}\tilde{m}}\,l},
	\end{align}
	\eqref{3.1} becomes:
	\begin{equation}
		\mathrm{d}s^2=\mathrm{d}\rho^2-\left(e^{\frac{\ 2\r}{\ l}}+e^{-\frac{\ 2\r}{\ l}}\right)\mathrm{d}x^{+}\mathrm{d}x^{-}+(\mathrm{d}x^{+})^2+(\mathrm{d}x^{-})^2.
	\end{equation}
	Here the two light-cone parameters are
	\begin{align}
		x^{\pm}=\sqrt{2\tilde{G_{3}}\tilde{m}}\,\tilde{t}\pm \sqrt{2\tilde{G_{3}}\tilde{m}}\,l\tilde{\phi},
	\end{align}
	with 
	\begin{align}
		x^{\pm}\sim x^{\pm} +2\pi l\sqrt{2\tilde{G_{3}}\tilde{m}}.
	\end{align}
	From \eqref{Banandos geometries}, we can also read off the constant $\mathcal{L}_{+}(x^{+})$ and $\mathcal{L}_{-}(x^{-})$ as $\frac{\ 1}{\ l^2}$. Accordingly, the function that parameterizes the boundary charges in \eqref{boundcond} can be expanded within a period as: 
	\begin{align}
		\tilde{\lambda}^{+}_{n}=\sqrt{2\tilde{G_{3}}\tilde{m}}\,l\, e^{\frac{\ -inx^{+}}{\ \sqrt{2\tilde{G_{3}}\tilde{m}}\,l}}, \quad \tilde{\lambda}^{-}_{n}=\sqrt{2\tilde{G_{3}}\tilde{m}}\,l\, e^{\frac{\ -inx^{-}}{\ \sqrt{2\tilde{G_{3}}\tilde{m}}\,l}}.
	\end{align}
	In this way, the zero mode of the generator is:
	\begin{align}
		\tilde{L}_{0}^{+}&=\frac{\ \tilde{k}}{\ 2\pi}\oint \tilde{\lambda}^{+}_{0}\mathcal{L}_{+}(x^{+})\sqrt{2\tilde{G_{3}}\tilde{m}}\,l\mathrm{d}\tilde{\phi}=\frac{\ \tilde{m}l}{\ 2}=\alpha L_{0}^{+},\\
		\tilde{L}_{0}^{-}&=\frac{\ \tilde{k}}{\ 2\pi}\oint \tilde{\lambda}^{-}_{0}\mathcal{L}_{-}(x^{-})\sqrt{2\tilde{G_{3}}\tilde{m}}\,l\mathrm{d}\tilde{\phi}=\frac{\ \tilde{m}l}{\ 2}=\alpha L_{0}^{-},
	\end{align}
	with the new CS level 
	\begin{align}
		\tilde{k}=\frac{\ \tilde{c}}{\ 6}=\frac{\ l}{\ 4\tilde{G_{3}}}=\frac{\ l}{\ 4\alpha G_{3}}.
	\end{align}
	Then we have:
	\begin{align}
		\frac{ \tilde{L}_{0}^{+}+\tilde{L}_{0}^{-}}{\ l}&=\tilde{m}=\alpha m=M_{\rm cBTZ},\\
		\tilde{L}_{0}^{+}-\tilde{L}_{0}^{-}&=0.
	\end{align}
	
	Next we consider the Cardy formula in this case.  The relation of the free energies between  the new and old theories is
	\begin{equation}
		\tilde{F}=\tilde{M}-\tilde{T}_{\rm h}\tilde{S}=\alpha (M-T_{\rm h}S)=\alpha F,
	\end{equation}
	with the $\tilde{S}=\alpha S$ the entropy of the sector and $S$ the entropy of a complete BTZ black hole. Thus the entropy of BTZ in the new theory is:
	\begin{equation}
		\tilde{S}_{\rm BH}=\alpha S_{\rm BTZ}=\alpha \beta M-\alpha \beta F=\alpha \tilde{\beta}\tilde{M}-\alpha\tilde{\beta} \tilde{F}.
	\end{equation}
	As a  consequence, the partition function, energy and entropy can be defined as:
	\begin{align}
		Z_{\rm cBTZ}&=e^{-\alpha\tilde{\beta}\tilde{F}},\label{TD1}\\
		\tilde{M}&=-\frac{1}{\alpha}\partial_{\tilde{\beta}}(\log Z_{\rm cBTZ}),\label{TD2}\\
		\tilde{S}_{\rm BH}&=\left(1-\tilde{\beta}\partial_{\tilde{\beta}}\right)\log Z_{\rm cBTZ}\label{TD3}.
	\end{align}
	Then considering the modular invariance, $Z(\tilde{\beta},\ \overline{\beta})=Z(-\frac{\ 4\pi^2 l^2}{\ \tilde{\beta}},\ -\frac{\ 4\pi^2 l^2}{\ \overline{\beta}})$, we set the free energy as 
	\begin{equation}
		\tilde{F}=-\frac{ \tilde{c}}{\ 24l}\left(\frac{\ 4\pi^2 l^2}{\ \tilde{\beta}^2}-1\right)-\frac{ \tilde{c}}{\ 24l}\left(\frac{\ 4\pi^2 l^2}{\ \tilde{\overline{\beta}^2}}-1\right),
	\end{equation}
	with $\tilde{\beta}$ and $\tilde{\overline{\beta}}$ the period of the imaginary time, corresponding to $\tilde{L}_{0}^{+}$ and $\tilde{L}_{0}^{-}$. Thus the partition function of the BTZ black hole in the new theory is:
	\begin{equation}\label{partion function}
		Z_{\rm cBTZ}=e^{\frac{\alpha \tilde{c}}{\ 24l}\left(\frac{\ 4\pi^2 l^2}{\ \tilde{\beta}}-\tilde{\beta}\right)+\frac{\alpha \tilde{c}}{\ 24l}\left(\frac{\ 4\pi^2 l^2}{\ \tilde{\overline{\beta}}}-\tilde{\overline{\beta}}\right)}.
	\end{equation}
	At high temperature ($\tilde{\beta}\rightarrow 0,\ \tilde{\overline{\beta}}\rightarrow 0$), the partition function in which we keep the classical term takes the form as:
	\begin{equation}
		Z_{\rm cBTZ}=e^{\frac{\alpha \pi^2 l \tilde{c}}{ 6\tilde{\beta}}+\frac{\alpha \pi^2 l\tilde{c}}{ 6\tilde{\overline{\beta}}}}.
	\end{equation}
	After applying \eqref{TD1}-\eqref{TD3}, we have:
	\begin{align}
		\tilde{L}_{0}^{+}&=-l\frac{\ \partial_{\tilde{\beta}}}{\ \alpha}\log Z_{\rm cBTZ}=\frac{\pi^2 l^2 \tilde{c}}{\ 6\tilde{\beta}^2}=\frac{ \tilde{m}l}{ 2},\label{Cardy formula1}\\
		\tilde{L}_{0}^{-}&=-l\frac{\ \partial_{\tilde{\overline{\beta}}}}{\ \alpha}\log Z_{\rm cBTZ}=\frac{\pi^2 l^2 \tilde{c}}{\ 6\tilde{\overline{\beta}}^2}=\frac{\tilde{m}l}{2},\label{Cardy formula2}\\
		S_{\rm BH}&=\left(1-\tilde{\beta}\partial_{\tilde{\beta}}-\tilde{\overline{\beta}}\partial_{\tilde{\overline{\beta}}}\right)\log Z_{\rm cBTZ}\nonumber\\
		&=\frac{\alpha\pi^2 l \tilde{c}}{\ 3\tilde{\beta}}+\frac{\alpha\pi^2 l \tilde{c}}{\ 3\tilde{\overline{\beta}}}=2\pi\alpha\sqrt{\frac{\ \tilde{c}}{\ 6}\tilde{L}_{0}^{+}}+2\pi\alpha\sqrt{\frac{\ \tilde{c}}{\ 6}\tilde{L}_{0}^{-}}\label{Cardy formula3},
	\end{align}
	then we get the modified Cardy formula for microstates in the new theory. As a result, given a large mass ($\tilde{m}\gg \tilde{c}$), by using \eqref{Cardy formula1}-\eqref{Cardy formula3} in the new theory, we have: 
	\begin{equation}\label{NCD}
		\tilde{S}_{\rm BH}=\frac{\ 2\pi\alpha}{\ 4G_{3}} 2\sqrt{2G_{3}m}\,l=\frac{\ \tilde{A}}{\ 4G_{3}},
	\end{equation}
	with $\tilde{A}$ the area of the horizon of the sector.
	Eventually, we can see that the result is in agreement with the law of area. Therefore, the entropy of the conical AdS can be obtained by counting the number of microstates of a new conformal field theory based on \eqref{NCD}.
	
\begin{figure}[h]
		\centering
		\includegraphics[width=0.70\linewidth]{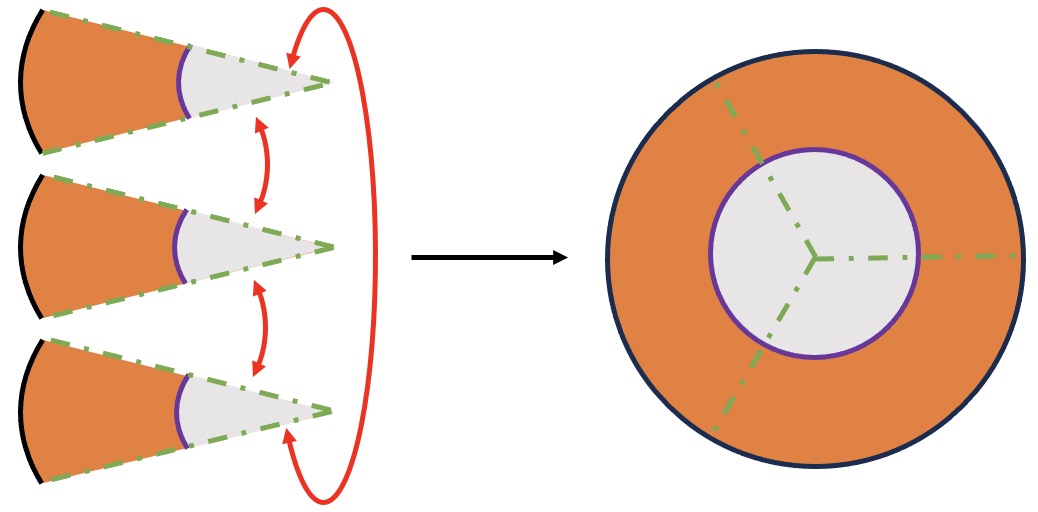}
		\caption{The process of lifting the sector (orange region) into the covering space for $n=3$. The sector is wrapped by branes (green dash-dot lines), horizon (purple curves) and asymptotic boundary (black curves). \textbf{Left}: Three copies of sector and the identification denoted by red arrows. \textbf{Right}: The covering space that is constructed by gluing three copies of sector together along the branes.}
		\label{covering space}
	\end{figure}

	At the end of this section, we make some comments on some special cases. When $\alpha=\frac{\ 1}{\ n}$ and $n\in \mathbb{Z}_{+}$, the old theory corresponds to a sector that has a defect in angular in comprison with $2\pi$ and the new one corresponds to the covering space in \cite{Balasubramanian:2014sra}. And the relation of the Virasoro algebra between these two theories is:
	\begin{align}
		\tilde{L}_{m}^{\pm}&=\alpha L_{m}^{\pm}, m\in\mathbb{Z}/\{0\},\\
		\tilde{L}_{0}^{\pm}-\frac{\ \tilde{c}}{\ 24}&=\alpha\left(L_{0}^{\pm}-\frac{\ c}{\ 24}\right),\\
		\tilde{c}&=\alpha c.
	\end{align}
	Therefore, as  said in \cite{Balasubramanian:2014sra}, the covering space is created by copying the sector $n$ times and then stitching these copies together along the branes, as  shown in Fig. $\ref{covering space}$. So the partition function of the new gravity theory in covering space is given by:
	\begin{equation}
		\tilde{Z}_{\rm BTZ}=Z_{\rm 1(cBTZ)}Z_{\rm 2(cBTZ)}...Z_{\rm n-1(cBTZ)}=Z_{\rm cBTZ}^{n},
	\end{equation}
	with $Z_{\rm i(cBTZ)}$ the partition function for $i_{\rm th}$ copy.
	Then the partition function for each copy is:
	\begin{equation}
		Z_{\rm cBTZ}=\sqrt[n]{\tilde{Z}_{\rm BTZ}},
	\end{equation}
	Thus we reinterpreted the relation in \eqref{Cardy formula3}. As we can see,  the information of the defect in angular  is restored by symmetrization, and the coefficient of the angular defect $\alpha$ in \eqref{TD1} can be understood as a projection from the covering space onto the defect geometry. The sector can be formally rewritten as the quotient group as: 
	\begin{equation}
		\rm AAdS_{\rm c}=\frac{\ \rm AAdS_{cs}}{\ Z_{n}},
	\end{equation}
	in which $\rm AAdS_{\rm cs} (\rm AAdS_{\rm c}) $ denotes AdS group of the covering space (sector). 
	
	At the end, we make some comments on the thermal AdS with a conical defect, which is interpreted as a particle. In this phase, the radial direction is not truncated by the horizon and the Hawking temperature is zero. As a result of low temperature ($\tilde{\beta}\rightarrow \infty,\ \tilde{\overline{\beta}}\rightarrow \infty$), the partition function in \eqref{partion function} at classical level takes the form as 
	\begin{equation}
		Z_{\rm cAdS}=e^{-\frac{\alpha \tilde{c}}{\ 24l}\tilde{\beta}-\frac{\alpha \tilde{c}}{\ 24l}\tilde{\overline{\beta}}}.
	\end{equation}
	In parallel, we apply \eqref{TD1}-\eqref{TD3}, and can get:
	\begin{align}
		\tilde{M}&=-\frac{\ \partial_{\tilde{\beta}}}{\ \alpha}\log Z_{\rm cAdS}-\frac{\ \partial_{\tilde{\overline{\beta}}}}{\ \alpha}\log Z_{\rm cAdS}=\frac{\ \tilde{c}}{12l}=\tilde{E}_{\rm vac},\\
		\tilde{S}&=\left(1-\tilde{\beta}\frac{\ \partial_{\tilde{\beta}}}{\ \alpha}-\tilde{\overline{\beta}}\frac{\ \partial_{\tilde{\overline{\beta}}}}{\ \alpha}\right)\log Z_{\rm cAdS}\equiv 0.
	\end{align}
	That is to say, for particles, the vaccum state dominates the canonical ensemble. The mass corresponds to Casimir energy in the new theory, and the entropy is always zero, which is consistent with the law of area.

	\section{Accelerating BTZ Black Hole in C-metric} \label{acbhc}
\begin{figure}[h]
		\centering
		\includegraphics[width=0.70\linewidth]{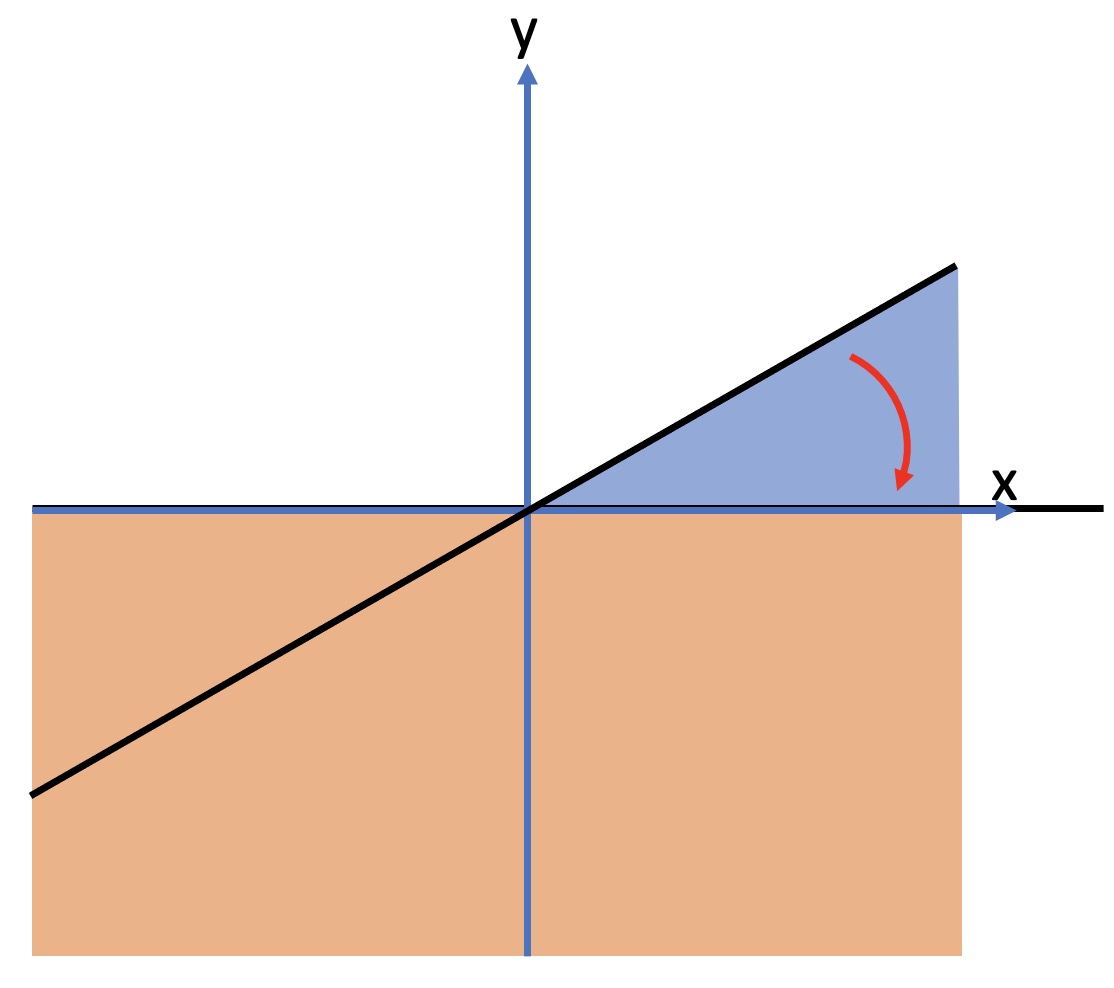}
		\caption{The position of the conformal boundary changes with the acceleration approaching zero. The dark lines are the conformal boundaries with different accelerations. The colored regions are the spaces of solutions correspondingly. The red arrow indicates the direction of the acceleration decreasing.}
		\label{boundary}
	\end{figure}

	We start with the line element  in \eqref{3dcmetric}. For convenience of investigating the solution in some limits of the acceleration, we take a transformation:
	\begin{equation}
		y\rightarrow \frac{y}{A}\, ,\quad \tau\rightarrow A\tau,
	\end{equation}
	then \eqref{3dcmetric} is rewritten as:
	\begin{equation}\label{3dCmetric2}
		\mathrm{d}s^2=\frac{1}{(Ax-y)^2}\left[-\tilde{P}(y)\mathrm{d}\tau^2+\frac{1}{\tilde{P}(y)}\mathrm{d}y^2+ \frac{1}{Q(x)}\mathrm{d}x^2\right],
	\end{equation}
	in which $\tilde{P}(y)$ and $Q(x)$ of three solutions are:
	\begin{equation}
		\begin{cases}\label{C-metric2}
			\text{Class I : } & \tilde{P}(y)=\frac{1}{l^2}+y^2-A^2 , \  Q(x)=1-x^2,\  \text{with}\  |x|\leq1,\\
			\\
			\text{Class II : } & \tilde{P}(y)=\frac{1}{l^2}+A^2 -y^2, \  Q(x)=x^2-1,\  \text{with}\  |x|\geq1,\\
			\\
			\text{Class III : } & \tilde{P}(y)=\frac{1}{l^2}-(A^2 +y^2), \  Q(x)=x^2+1,\  \text{with}\  x\in\mathbb{R}.
		\end{cases}
	\end{equation}
	Considering that the region of $x$ is bounded \cite{previouswork}, when the acceleration  approaches zero, $Ax$ in the conformal factor also approaches zero. Since the confomal boundary is $\frac{y}{x}=A$, with $A$ approaching zero  the conformal boundary will approach $y=0$. Fig.~\ref{boundary} shows the change of boundary with $A$ approaching zero. When $A$ is zero, Class II and Class III turn into BTZ black holes, nevertheless Class I turns into a particle with no horzion. In the following we'll discuss the thermodynamics and the holographic RG flow.
	
	\subsection{Thermodynamics of C-metric} \label{thermocmetric}
	We consider the Fefferman-Graham (FG) expansion of \eqref{3dCmetric2} near the boundary to decouple $x$ and $y$. We perform a transformation using an infinite-ordered polynomial as follows: 
	\begin{equation}\label{Transformation}
		y=A\xi+\sum^{\infty}_{m=1}AF_{m}(\xi)\left(\frac{z}{l}\right)^{m}, \quad x=\xi+ \sum^{\infty}_{m=1}G_{m}(\xi)\left(\frac{z}{l}\right)^{m}.
	\end{equation}
	In this transformation, $z$ is the distance from the boundary $(x=y)$ and $\xi$ denotes the angular direction near the boundary. We solve $F_{m}$ and $G_{m}$ order by order and finally express them in terms of
	\begin{equation}
		F_{1}=\frac{(1-A^2l^2Q(\xi))^{\frac{3}{2}}}{Al \omega(\xi)},
	\end{equation}
	where $Q(\xi)$ is the factor in \eqref{3dcmetric} and $\omega(\xi)$ is an undetermined  gauge factor. As shown in another related work \cite{previouswork}, not all asymptotic algebras with arbitrary gauge can form the Virasoro algebra. For simplification, we set $\omega(\xi)=1$\cite{arenas2022acceleration}. After taking \eqref{Transformation}, the metric of these three solutions take a form as
	\begin{equation}
		\mathrm{d} s^2=\frac{l^2}{z^2}\mathrm{d} z^2 + \frac{l^2}{z^2}g_{ij}(\tilde{x},z)\mathrm{d} \tilde{x}^{i}\mathrm{d} \tilde{x}^{j},
	\end{equation}
	with 
	\begin{equation}
		g_{ij}(x,z)=g_{(0)}+g_{(2)}z^2+h_{(2)}z^2 \log(z^2)+\mathcal{O}(z^3).
	\end{equation}
	The stress-energy tensor can be derived in \cite{skenderis2000quantum} as:
	\begin{equation}\label{holo-mass}
		T[\gamma]=\frac{l}{8\pi G_{3}}\left(g_{(2)}-g_{(0)}\text{Tr}[g_{(0)}^{-1}g_{(2)}]\right),
	\end{equation}
	and the mass is defined as:
	\begin{equation}\label{holo-mass2}
		M=\int^{x=x_{2}}_{x=x_{1}} \mathrm{d}\xi \sqrt{-g_{(0)}}T^{\tau}_{\tau(\text{ren})}.
	\end{equation}
	
	In semi-classical approximation, the relation between the on-shell action  and the free energy $F$ in a black hole is\cite{tian2024aspects}:
	\begin{equation}\label{free energy}
		I_{\rm on-shell}=I_{\rm EH}+I_{\rm GHY}+I_{\rm ct}+I_{\rm brane}=\beta F=\beta M-S,
	\end{equation}
	with $\beta=\frac{1}{T_{\rm h}}$. In \cite{previouswork}, the Smarr relation is formally represented as:
	\begin{equation}\label{Smarr Relation}
		2M=T_{\rm h}(S-S_{\rm boundary}).
	\end{equation}
	Combining \eqref{free energy} and \eqref{Smarr Relation}, we have:
	\begin{equation}\label{on-shell action}
		I_{\rm on-shell}=-(\beta M+S_{\rm boundary}).
	\end{equation}
	In the following we consider these phases in C-metric which describe accelerating BTZ black holes, they are denoted as I$_{\rm b1}$, II$_{\rm b1}$, II$_{\rm b2}$ and III$_{\rm b}$ in \cite{previouswork} and contain two branes at $x=x_{1}$ (domain-wall) and $x=x_{2}$ (strut-wall) ($x_{1}<x_{2}$). The brief introduction and the corresponding computation of these four phases are shown below.

\begin{figure}[h]
		\centering
		\includegraphics[width=0.70\linewidth]{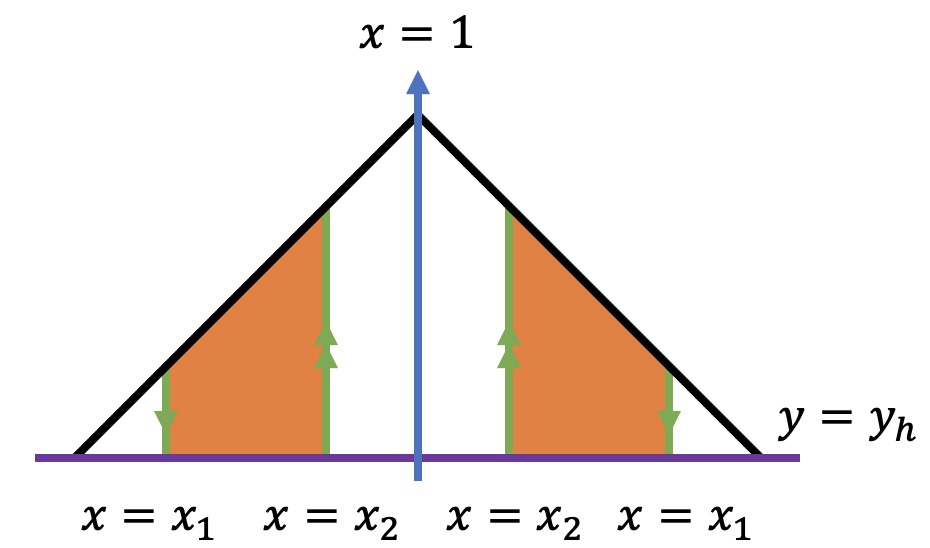}
		\caption{The two copies of the sector (orange region) of I$_{\rm b1}$ in ($x,y$). The sector is wrapped by two branes at $x=x_{1}$ (domain-wall) and $x=x_{2}$ (strut-wall) (green lines), horizon (purple line) and asymptotic boundary (black curves).}
		\label{Class I(one brane)}
	\end{figure}

	\textbf{\uppercase\expandafter{\romannumeral1}$_{\rm b1}$}:
	The metric of the accelerating BTZ black hole in Class \uppercase\expandafter{\romannumeral1} is shown in the first line of \eqref{C-metric} with $Al>1$, $y> y_{\rm h_1}$ and $y_{\rm h_1}=\frac{\ \sqrt{A^2l^2-1}}{\ Al}<x_{1}<\xi<x_{2}\leq1$, as illustrated in Fig. $\ref{Class I(one brane)}$.
	By using the transformation \eqref{Transformation}, the line element  is rewritten in FG as:
	\begin{equation}\label{FG in Class I}
		\mathrm{d}s^2=-\frac{\ \left(z^2+l^2(4-A^2 z^2)\right)^2}{\ 16  l^2 z^2}\mathrm{d}\tau^2+\frac{\ l^2}{\ z^2}\mathrm{d}z^2+\frac{\ l^2\left(z^2-l^2(4+A^2 z^2)\right)^2}{\ 16z^2(1-\xi^2)(1-A^2 l^2(1-\xi^2))^2}\mathrm{d}\xi^2.
	\end{equation}
	To use the previous conclusion in Section \ref{btzbh} directly, we make a transformation:
	\begin{equation}
		z=\frac{\ 2le^{\frac{\ \r}{\ l}}}{\ \alpha\sqrt{A^2l^2-1}}, \quad \tau=\frac{\ 2\alpha (x^{+}+x^{-})}{\ \sqrt{A^2l^2-1}}, \quad \Xi=f(\xi)-f(x_{2})=\frac{\ 2\alpha (x^{+}-x^{-})}{\ \sqrt{A^2l^2-1}},
	\end{equation}
	where
	\begin{equation}
		f(\xi)=-\frac{l \;\text{arctanh}\left(\sqrt{A^2l^2-1}\sqrt{\frac{1}{\xi^2}-1}\right)}{\sqrt{A^2 l^2-1}} \, , \quad  \text{for} \  \text{I}_{\rm b1},
	\end{equation}
	with $\Xi\in[0,\pi\alpha l]$. Then by using \eqref{holo-mass2}, the holographic mass is 
	\begin{equation}\label{Mass1}
		M_{\rm I_{b1}}=\L.\frac{\ \sqrt{A^2l^2-1}\,\mathrm{arctanh}\left(\frac{\ \sqrt{A^2l^2-1}\sqrt{1-x^2}}{\ x}\right)}{\ 8\pi G_{3}}\R|^{x=x_{1}}_{x=x_{2}}, 
	\end{equation}
	and the Hawking temperature $T_{\rm h_1}$ is $\frac{\sqrt{A^2l^2-1}}{2\pi l}$ \cite{tian2024aspects}.
	Then the metric becomes:
	\begin{equation}\label{C1}
		\mathrm{d}s^2=\mathrm{d}\r^2-\left(e^{\frac{\ 2\r}{\ l}}+\alpha^4 e^{-\frac{\ 2\r}{\ l}}\right)\mathrm{d}x^{+}\mathrm{d}x^{-}+\alpha^2(\mathrm{d}x^{+})^2+\alpha^2(\mathrm{d}x^{-})^2
	\end{equation}
	where $\alpha=\frac{\ \Xi |_{\xi=x_{1}}}{\ \pi l}$. The boundary entropy of  brane is shown in \cite{previouswork} as:
	\begin{equation}
		S_{\rm boundary}=  \frac{l}{2G_{3}}\L.\mathrm{arctanh}\left(Al\sqrt{1-x^2}\right)\R|^{x=x_{1}}_{x=x_{2}}.
	\end{equation}
	When $x_{1}<x_{2}<0$, the boundary entropy $S_{\rm boundary}<0$, when $0<x_{1}<x_{2}$, the boundary entropy $S_{\rm boundary}>0$.
	Based on \eqref{on-shell action}, we have:
	\begin{equation}
		\begin{aligned}\label{onshell action of Class I}
			I_{\rm on-shell}&=-\left(\L.\frac{l}{4G_{3}}\mathrm{arctanh}\left(\frac{\ \sqrt{A^2l^2-1}\sqrt{1-x^2}}{\ x}\right)\R|^{x=x_{1}}_{x=x_{2}}+S_{\rm boundary}\right),\\
			&=-\left(\frac{\pi^2l^2}{2G_{3}}\frac{\alpha}{\beta}+S_{\rm boundary}\right),\\
			&=-\frac{\pi^2lc}{3}\frac{\alpha}{\beta}-S_{\rm boudanry},
		\end{aligned}
	\end{equation}
	with $c=\frac{3l}{2G_{3}}$ the central charge of AdS$_{3}$ and $\beta=\frac{1}{T_{\rm h_1}}$.
	
\begin{figure}[h]
		\centering
		\includegraphics[width=0.40\linewidth]{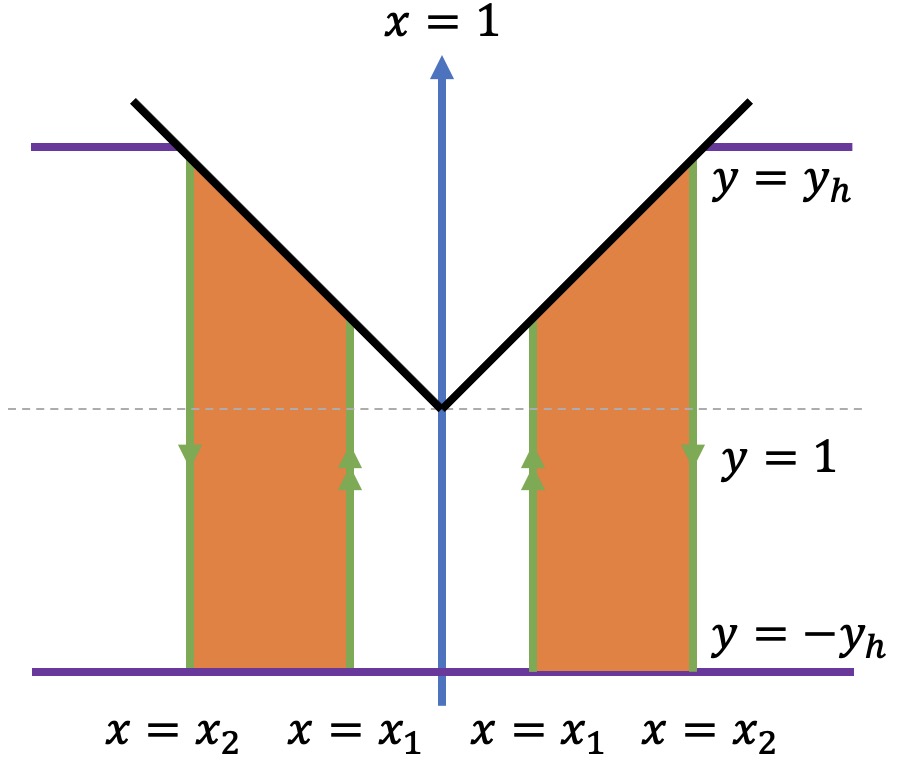}
		\includegraphics[width=0.45\linewidth]{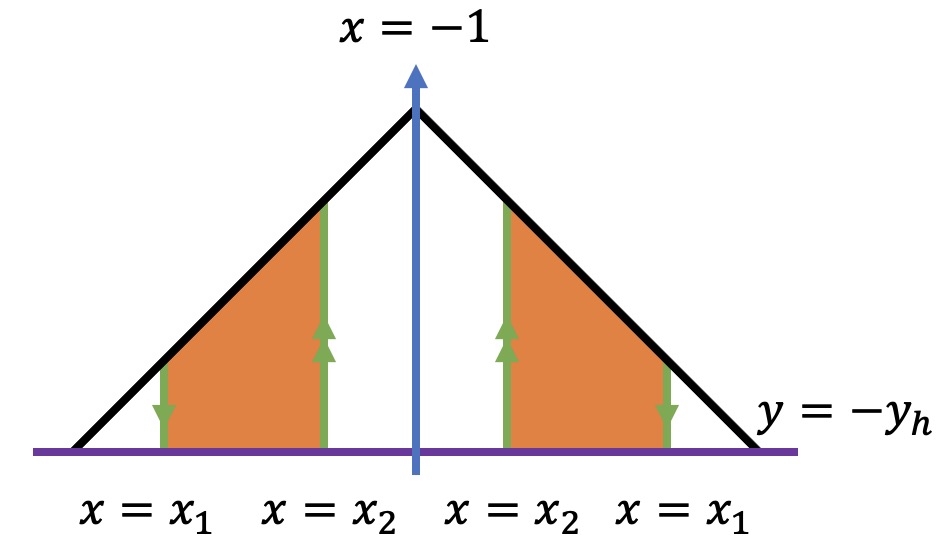}
		\caption{The two copies of the sector (orange region) of accelerating BTZ black hole of Class II solution in ($x,y$). \textbf{Left}: II$_{\rm b2}$, $1\leq x_{1}<\xi<x_{2}<\frac{\ \sqrt{1+A^2l^2}}{\ Al}=y_{\rm h_2}$. \textbf{Right}: II$_{\rm b1}$, $-y_{\rm h_2}=-\frac{\ \sqrt{1+A^2l^2}}{\ Al}<x_{1}<\xi<x_{2}\leq -1$. The sector is wrapped by two branes at $x=x_{1}$ (domain-wall) and $x=x_{2}$ (strut-wall) (green lines), horizon (purple line) and asymptotic boundary (black curves).}
		\label{Class II(one brane)}
	\end{figure}

	\textbf{Class \uppercase\expandafter{\romannumeral2}}:
	The metric of the accelerating BTZ black hole in Class \uppercase\expandafter{\romannumeral2} is shown in the second line of \eqref{C-metric} with $1\leq x_{1}<\xi<x_{2}<y_{\rm h_2}=\frac{\ \sqrt{1+A^2l^2}}{\ Al}$ or $-\frac{\ \sqrt{1+A^2l^2}}{\ Al}=-y_{\rm h_2}<x_{1}<\xi<x_{2}\leq -1$, as  illustrated in Fig. $\ref{Class II(one brane)}$.
	By using the transformation \eqref{Transformation}, the line element is rewritten in FG as:
	\begin{equation}\label{FG in Class II}
		\mathrm{d}s^2=-\frac{\ (z^2+l^2(A^2z^2-4))^2}{16  l^2 z^2}\mathrm{d}\tau^2 +\frac{\ l^2}{\ z^2}\mathrm{d}z^2+\frac{\ (z^2+l^2(4+A^2 z^2))^2}{\ 16z^2(\xi^2-1)(-1+A^2l^2(-1+\xi^2))^2}\mathrm{d}\xi^2 .
	\end{equation} 
	Similar to Class I, we make a transformation: 
	\begin{equation}
		z=\frac{\ 2le^{\frac{\ \r}{\ l}}}{\ \alpha\sqrt{1+A^2l^2}}, \quad \tau=\frac{\ 2\alpha (x^{+}+x^{-})}{\ \sqrt{1+A^2l^2}}, \quad \Xi=f(\xi)-f(x_{2})=\frac{\ 2\alpha (x^{+}-x^{-})}{\ \sqrt{1+A^2l^2}},
	\end{equation}
	where
	\begin{equation}
		f(\xi):=
		\begin{cases}
			-\frac{\ l\,\rm{arctanh}\left(\frac{\ \sqrt{A^2l^2+1}\sqrt{\xi^2-1}}{\ \xi}\right)}{\ \sqrt{A^2 l^2+1}} & \text{for} \  \text{II}_{\rm b2}, \\
			\\
			\frac{\ l\,\rm{arctanh}\left(\frac{\ \sqrt{A^2l^2+1}\sqrt{\xi^2-1}}{\ \xi}\right)}{\ \sqrt{A^2 l^2+1}} & \text{for} \  \text{II}_{\rm b1}, 
		\end{cases}
	\end{equation}
	with $\Xi\in[0,\pi\alpha l]$. Then by using \eqref{holo-mass2}, the masses of II$_{\rm b1}$ and II$_{\rm b2}$ are both: 
	\begin{equation}\label{Mass2}
		M_{\rm II_b}=\L.\frac{\ \sqrt{A^2l^2+1}\,\mathrm{arctanh}\left(\frac{\ \sqrt{A^2l^2+1}\sqrt{x^2-1}}{\ x}\right)}{\ 8\pi G_{3}}\R|^{x=x_{2}}_{x=x_{1}}, 
	\end{equation}
	and the temperature is $T_{\rm h_2}=\frac{\sqrt{A^2l^2+1}}{2\pi l}$ \cite{tian2024aspects}. Then metric becomes:
	\begin{equation}\label{C2}
		\mathrm{d}s^2=\mathrm{d}\r^2-\left(e^{\frac{\ 2\r}{\ l}}+\alpha^4 e^{-\frac{\ 2\r}{\ l}}\right)\mathrm{d}x^{+}\mathrm{d}x^{-}+\alpha^2(\mathrm{d}x^{+})^2+\alpha^2(\mathrm{d}x^{-})^2,
	\end{equation}
	where $\alpha=\frac{\ |\Xi |_{\xi=\pm x_{0}}|}{\ \pi l}$. The boundary entropies of two branes in II$_{\rm b1}$ and II$_{\rm b2}$ are both:
	\begin{equation}
		S_{\rm boundary}=
		\frac{l}{2G_{3}}\L.\mathrm{arctanh}\left(Al\sqrt{x^2-1}\right)\R|^{x=x_{1}}_{x=x_{2}}. 
	\end{equation}
	When $x_{1}<x_{2}<-1$, $S_{\rm boundary}>0$, when $1<x_{1}<x_{2}$, $S_{\rm boundary}<0$.
	In parallel, based on \eqref{on-shell action}, we have:
	\begin{equation}
		\begin{aligned}\label{onshell action of Class II}
			I_{\rm on-shell}&=-\left(\frac{\ l}{\ 4G_{3}}\L.\mathrm{arctanh}\left(\frac{\ \sqrt{A^2l^2-1}\sqrt{1-x^2}}{\ x}\right)\R|^{x=x_{1}}_{x=x_{2}}+S_{\rm boundary}\right),\\
			&=-\frac{\pi^2lc}{3}\frac{\alpha}{\beta}-S_{\rm boundary},
		\end{aligned}
	\end{equation}
	with $c=\frac{3l}{2G_{3}}$ the central charge of AdS$_{3}$ and $\beta=\frac{1}{T_{\rm h_2}}$.
	
\begin{figure}[h]
		\centering
		\includegraphics[width=0.50\linewidth]{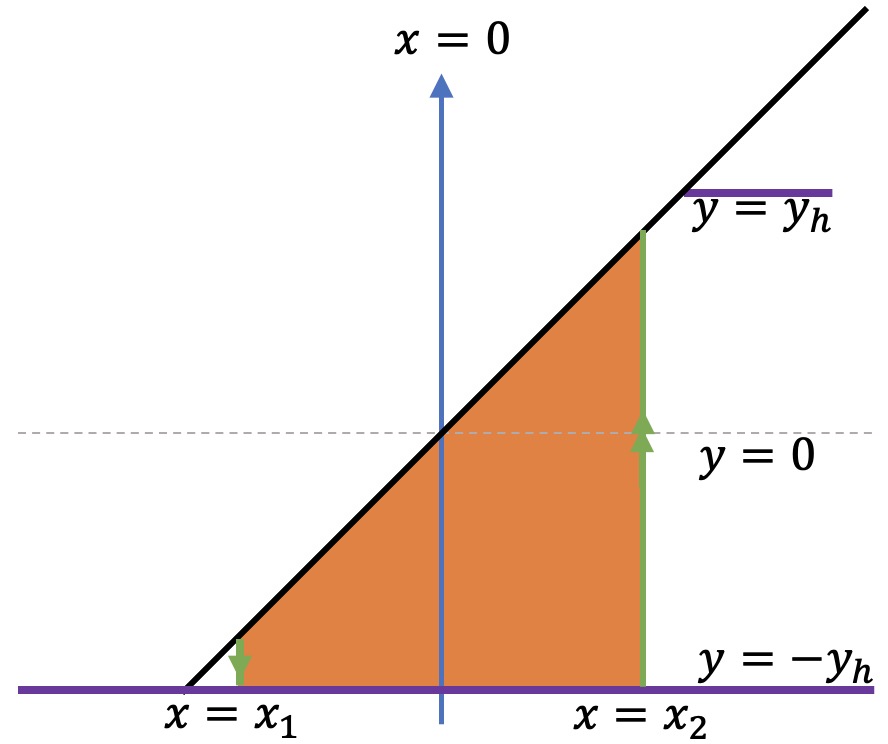}
		\caption{The sector (orange region) of the accelerating BTZ black hole of Class III solution in ($x,y$). $-y_{\rm h_3}<x_1<x_2<y_{\rm h_3}=\frac{\sqrt{1-A^2l^2}}{Al}$. The sector is wrapped by two branes at $x=x_{1}$ (domain-wall) and $x=x_{2}$ (strut-wall) (green lines), horizon (purple line) and asymptotic boundary (black curves).}
		\label{Class III(two brane)}
	\end{figure}

	\textbf{Class \uppercase\expandafter{\romannumeral3}}:
	The metric of the accelerating BTZ black hole in Class \uppercase\expandafter{\romannumeral3} is shown in the third line of \eqref{C-metric} with $Al<1$ and $-\frac{\sqrt{1-A^2l^2}}{Al}=-y_{\rm h_3}<x_1<\xi<x_2< y_{\rm h_3}=\frac{\sqrt{1-A^2l^2}}{Al}$, as is illustrated in Fig. $\ref{Class III(two brane)}$.
	By using the transformation \eqref{Transformation}, the  line element is rewritten in FG as:
	\begin{equation}\label{FG in Class III}
		\mathrm{d}s^2=-\frac{\ \left(z^2-l^2(4+A^2 z^2)\right)^2}{\ 16  l^2 z^2}\mathrm{d}\tau^2+\frac{\ l^2}{\ z^2}\mathrm{d}z^2+\frac{\ \left(z^2+l^2(4-A^2z^2)\right)^2}{\ 16z^2(1+\xi^2)(-1+A^2l^2(1+\xi^2))^2}\mathrm{d}\xi^2 .
	\end{equation}
	In parallel, we make a transformation:
	\begin{equation}
		z=\frac{\ 2le^{\frac{\ \r}{\ l}}}{\ \alpha\sqrt{1-A^2l^2}}, \quad \tau=\frac{\ 2\alpha (x^{+}+x^{-})}{\ \sqrt{1-A^2l^2}}, \quad \Xi=f(x_{2})-f(\xi)=\frac{\ 2\alpha (x^{+}-x^{-})}{\ \sqrt{1-A^2l^2}},
	\end{equation}
	in which $\xi \in [x_{1},x_{2}]$ and
	\begin{equation}
		f(\xi)=\frac{\ l \, \mathrm{arctanh}\left(\frac{\ \xi}{\ \sqrt{1-A^2l^2}\sqrt{1+\xi^2}}\right)}{\ \sqrt{1-A^2 l^2}} \quad   \text{for} \  \text{III}_{\rm b}, 
	\end{equation}
	with $\Xi\in[0,\pi\alpha l]$. Then by using \eqref{holo-mass2}, the mass is 
	\begin{equation}\label{Mass3}
		M_{\rm III_b}=\L.\frac{\ \sqrt{1-A^2l^2}\,\mathrm{arctanh}\left(\frac{\ x}{\sqrt{1-A^2l^2}\sqrt{1+x^2}}\right)}{\ 8\pi G_{3}}\R|^{x=x_{2}}_{x=x_{1}}, 
	\end{equation}
	and the temperature is $T_{\rm h_3}=\frac{\sqrt{1-A^2l^2}}{2\pi l}$ \cite{tian2024aspects}. Then the metric becomes:
	\begin{equation}\label{C3}
		\mathrm{d}s^2=\mathrm{d}\r^2-\left(e^{\frac{\ 2\r}{\ l}}+\alpha^4 e^{-\frac{\ 2\r}{\ l}}\right)\mathrm{d}x^{+}\mathrm{d}x^{-}+\alpha^2(\mathrm{d}x^{+})^2+\alpha^2(\mathrm{d}x^{-})^2,
	\end{equation}
	where $\alpha=\frac{\ \Xi |_{\xi=x_{1}}}{\ \pi l}$. The boundary entropy of two branes is:
	\begin{equation}
		S_{\rm boundary}=\frac{l}{2G_{3}}\L.\mathrm{arctanh}\left(Al\sqrt{1+x^2}\right)\R|^{x=x_{1}}_{x=x_{2}}.
	\end{equation}
	When $|x_{1}|>|x_{2}|$, $S_{\rm boundary}>0$, when $|x_{1}|<|x_{2}|$, $S_{\rm boundary}<0$.
	Also based on \eqref{on-shell action}, we have:
	\begin{equation}
		\begin{aligned}\label{onshel action of Class III}
			I_{\rm on-shell}&=-\left(\L.\frac{l}{4G_{3}}\mathrm{arctanh}\left(\frac{x}{\sqrt{1-A^2l^2}\sqrt{1+x^2}}\right)\R|^{x=x_{2}}_{x=x_{1}}+S_{\rm boundary}\right)\\
			&=-\frac{\pi^2 l c}{3}\frac{\alpha}{\beta}-S_{\rm boundary},
		\end{aligned}
	\end{equation}
	with $c=\frac{3l}{2G_{3}}$ the central charge of AdS$_{3}$ and $\beta=\frac{1}{T_{\rm h_3}}$.
	
	As we can see above, \eqref{C1}, \eqref{C2} and \eqref{C3} have the same form and can be summarized as:
	\begin{equation}
		ds_{i}^2=\mathrm{d}\r^2-\left(\alpha_{i}^4e^{-\frac{\ 2\r}{\ l}}+e^{\frac{\ 2\r}{\ l}}\right)\mathrm{d}x^{+}\mathrm{d}x^{-}+\alpha_{i}^2(\mathrm{d}x^{+})^2+\alpha_{i}^2(\mathrm{d}x^{-})^2,\quad i=1,2,3
	\end{equation}
	with $\alpha_{i}$ defined as
	\begin{equation}\label{conical angular in C-metric}
		\alpha_{i}:=
		\begin{cases}
			\L.\frac{ \mathrm{arctanh}\left(\frac{ \sqrt{A^2l^2-1}\sqrt{1-x^2}}{\ x}\right)}{ \sqrt{A^2 l^2-1}\pi}\R|^{x=x_{1}}_{x=x_{2}} & i=1, \; \rm for \; I_{\rm b1}, \\
			\\
			\L.\frac{ \mathrm{arctanh}\left(\frac{\ \sqrt{A^2l^2+1}\sqrt{x^2-1}}{\ x}\right)}{\sqrt{A^2 l^2+1}\pi}\R|^{x=x_{2}}_{x=x_{1}} & i=2, \; \rm for \; II_{\rm b1}\; and\; II_{\rm b2}, \\
			\\
			\L.\frac{ \mathrm{arctanh}\left(\frac{\ x}{\ \sqrt{1-A^2l^2}\sqrt{1+x^2}}\right)}{\sqrt{1-A^2 l^2}\pi}\R|^{x=x_{2}}_{x=x_{1}} & i=3, \; \rm for \; III_{\rm b}.
		\end{cases}
	\end{equation}
	Accordingly, the on-shell actions \eqref{onshell action of Class I}, \eqref{onshell action of Class II} and \eqref{onshel action of Class III} can be summed up as:
	\begin{equation}\label{on-shell actions of all black hole phases}
		I_{\text{on-shell}(i)}=-\frac{\pi^2 l c_{i}}{3}\frac{\alpha_{i}}{\beta_{i}}-S_{\text{boundary}(i)}.
	\end{equation}
	Furthermore, without rotation, \eqref{on-shell actions of all black hole phases} can be written as:
	\begin{equation}\label{on-shell actions of all black hole phases2}
		I_{\text{on-shell}(i)}=-\frac{\pi^2 l c_{i}}{6}\frac{\alpha_{i}}{\beta_{i}}-\frac{\pi^2 l c_{i}}{6}\frac{\alpha_{i}}{\overline{\beta_{i}}}-S_{\text{boundary}(i)},
	\end{equation}
	with $\beta_{i}$ and $\overline{\beta_{i}}$ the period of right chiral mode and left chiral mode, and $\beta_{i}=\overline{\beta_{i}}$. Unlike the typical high-temperature regime observed in CFT$_{2}$, there exists an extra term $S_{\text{boundary}(i)}$, which comes from the coupling field. Although we don't  exactly know what the quantum field theory of the boundary is, we can obtain the on-shell action of the boundary theory according to the AdS/CFT correspondence. 
	
	Here we promote the coefficient $\alpha_{i}$ of the angular defect that we have introduced in Section \ref{btzbh} to be an arbitrary real number. From \eqref{Banandos geometries}, we can read off the constants $\mathcal{L}_{+}(x^{+})$ and $\mathcal{L}_{-}(x^{-})$ both as $\frac{\ \alpha_{i}^2}{\ l^2}$. As shown in \cite{arenas2022acceleration}, we glue two copies of the patch that is cut from the entire manifold  by two branes. Then the period of the light-cone coordinate, $x^{\pm}$, is $\tau_{i}$, which meets:
	\begin{equation}\label{Mass4}
		x_{i}^{\pm}\sim x_{i}^{\pm}+2\tau_{i}\pi l, \quad \tau_{i}:=
		\begin{cases}
			\frac{\ \sqrt{A^2l^2-1}}{\ 2} & i=1, \; \rm for \; I_{\rm b1}, \\
			\\
			\frac{\ \sqrt{A^2l^2+1}}{\ 2} & i=2, \; \rm for \; II_{\rm b1}\; and\; II_{\rm b2}, \\
			\\
			\frac{\ \sqrt{1-A^2l^2}}{\ 2} & i=3, \; \rm for \; III_{\rm b}.
		\end{cases}   
	\end{equation}
	The boundary condition is parameterized by two functions: 
	\begin{equation}
		\tilde{\lambda}^{+}_{n(i)}=\tau_{i}le^{\frac{\ -inx^{+}}{\ \tau_{i}l}}, \quad \tilde{\lambda}^{-}_{n(i)}=\tau_{i}l e^{\frac{\ -inx^{-}}{\ \tau_{i}l}}, \quad i=1,2,3,
	\end{equation}
	and the angle is
	\begin{equation}
		\frac{\ x^{+}}{\ l\tau_{i}}\sim \frac{\ x^{+}}{\ l\tau_{i}}+2\pi,\quad i=1,2,3.
	\end{equation}
	To facilitate writing integrals, we define the $x^{+}=t+l\phi$. At $t=0$, we have:
	\begin{equation}
		\phi\sim\phi+2\tau_{i}\pi,\quad i=1,2,3.
	\end{equation}
	
	To show the relation between the bulk gravity and the boundary CFT, we calculate the mass and the entropy by the asymptotic algebra in the new gravity theory by following Section \ref{btzbh}. In the new theory, we set: 
	\begin{equation}
		\tilde{G}_{3(i)}=\alpha_{i}G_{3}, \quad \tilde{c}_{i}=\frac{\ l}{\ \tilde{G}_{3(i)}}, \quad \tilde{k}_{i}=\frac{\ \tilde{c}_{i}}{\ 6}, \quad i=1,2,3.
	\end{equation}
	Then by following the previous process, we take an integration  of  \eqref{0 mode2} at $\tau=0$:
	\begin{equation}\label{zero mode in new theroy}
		\tilde{L}_{0(i)}^{\pm}=\frac{\ \tilde{k}_{i}\tau_{i}l}{\ 2\pi}\oint \tilde{\lambda}^{\pm}_{0(i)}\mathcal{L}_{\pm(i)}(x^{\pm}) \mathrm{d}\phi,
	\end{equation}
	and the spectrum of mass is given by:
	\begin{equation}
		\begin{aligned}\label{M for C}
			M_{i}=\frac{\ \tilde{L}^{+}_{0(i)}+\tilde{L}^{-}_{0(i)}}{\ l}=
			\begin{cases}
				\L.\frac{ \sqrt{A^2l^2-1}\,\mathrm{arctanh}\left(\frac{\ \sqrt{A^2l^2-1}\sqrt{1-x^2}}{\ x}\right)}{ 8\pi G_{3}}\R|^{x=x_{1}}_{x=x_{2}} & i=1, \; \rm for \; I_{\rm b1}, \\
				\\
				\L.\frac{ \sqrt{A^2l^2+1}\,\mathrm{arctanh}\left(\frac{\ \sqrt{A^2l^2+1}\sqrt{x^2-1}}{\ x}\right)}{ 8\pi G_{3}}\R|^{x=x_{2}}_{x=x_{1}} & i=2, \; \rm for \; II_{\rm b1}\; and\; II_{\rm b2}, \\
				\\
				\L.\frac{ \sqrt{1-A^2l^2}\,\mathrm{arctanh}\left(\frac{\ x}{\ \sqrt{1-A^2l^2}\sqrt{1+x^2}}\right)}{ 8\pi G_{3}}\R|^{x=x_{2}}_{x=x_{1}} & i=3, \; \rm for \; III_{\rm b},
			\end{cases}
		\end{aligned}
	\end{equation}
	which is consistent with \eqref{Mass1}, \eqref{Mass2} and \eqref{Mass3}. Comparing \eqref{conical angular in C-metric},\eqref{Mass4} and \eqref{M for C}, the C-metric can be equivalently seen as a sector with the angular defect $\alpha_{i}$, which is cut out from a BTZ with mass $\frac{\ \tau_{i}^2}{\ 2G_{3}}$, as we embed the metric into the Rindler geometry and introduce a rescaling $\alpha$ of time in \cite{arenas2022acceleration}. Accordingly, the temperature in the new theory is: 
	\begin{equation}
		\tilde{T}_{h(i)}=\frac{\ 1}{\ \tilde{\beta}_{i}}=\frac{\ \tau_{i}\alpha_{i}}{\ \pi l}=\frac{\alpha_{i}}{\beta_{i}},
	\end{equation}
	and the on-shell action \eqref{on-shell actions of all black hole phases2} becomes:
	\begin{equation}
		\begin{aligned}\label{onshell action in new theory}
			I_{\text{on-shell}(i)}&=-\frac{\pi^2 l c_{i}}{6}\frac{\alpha_{i}}{\beta_{i}}-\frac{\pi^2 l c_{i}}{6}\frac{\alpha_{i}}{\overline{\beta_{i}}}-S_{\text{boundary}(i)},\\
			&=-\frac{\pi^2 l \tilde{c}_{i}}{6}\frac{\alpha_{i}}{\tilde{\beta}_{i}}-\frac{\pi^2 l \tilde{c}_{i}}{6}\frac{\alpha_{i}}{\tilde{\overline{\beta}}_{i}}-\alpha_{i} \tilde{S}_{\text{boundary}(i)},\quad i=1,2,3,
		\end{aligned}
	\end{equation}
	with $\tilde{S}_{\text{boundary}(i)}=\frac{S_{\text{boundary}(i)}}{\alpha_{i}}$. Then the partition function is given as:
	\begin{equation}\label{partition function}
		Z_{i}=e^{-I_{\text{on-shell}(i)}}=e^{\frac{\pi^2 l \tilde{c}_{i}}{6}\frac{\alpha_{i}}{\tilde{\beta}_{i}}+\frac{\pi^2 l \tilde{c}_{i}}{6}\frac{\alpha_{i}}{\tilde{\overline{\beta}}_{i}}+\alpha_{i} \tilde{S}_{\text{boundary}(i)}},\quad i=1,2,3.
	\end{equation}
	Given \eqref{partion function}, we can further get:
	\begin{equation}
		S_{BH}=(1-\tilde{\beta}\p_{\tilde{\beta}}-\tilde{\overline{\beta}}\p_{\tilde{\overline{\beta}}})Z_{i}=\alpha_{i}\left(\frac{\pi^2l\tilde{c}}{3\tilde{\beta}_{i}}+\frac{\pi^2l\tilde{c}}{3\tilde{\overline{\beta}}_{i}}+\tilde{S}_{\text{boundary}(i)}\right),\quad i=1,2,3,
	\end{equation}
	and the zero mode of Virasoro algebra is given by \eqref{zero mode in new theroy}:
	\begin{equation}
		\label{Zero mode in new theory of C-metric}
		\tilde{L}^{+}_{0(i)}=\frac{\pi^2l^2\tilde{c}}{6\tilde{\beta}^2_{i}},\quad 
		\tilde{L}^{-}_{0(i)}=\frac{\pi^2l^2\tilde{c}}{6\tilde{\overline{\beta}}^2_{i}}.
	\end{equation}
	
	As shown in \eqref{M for C}, for I$_{\rm b1}$ and II$_{\rm b1}$,II$_{\rm b2}$, the large mass can be obtained by  a large acceleration $A$ or  a large $\alpha_{i}$, and for III$_{\rm b}$, the large mass can only be obtained by a large $\alpha_{i}$.
	Then after applying the result in \eqref{onshell action in new theory}-\eqref{Zero mode in new theory of C-metric} under conditions of large masses ($M_{i}\gg \tilde{c}_{i}$), we can get:
	\begin{align}
		S_{\rm BH}&=S_{i}+S_{\text{boundary}(i)}\\
		&=\alpha_{i}\left(2\pi \sqrt{\frac{\ \tilde{c}_{i}\tilde{L}^{+}_{0(i)}}{\ 6}}+2\pi \sqrt{\frac{\ \tilde{c}_{i}\tilde{L}^{-}_{0(i)}}{\ 6}}+\tilde{S}_{\text{boundary}(i)}\right)\\
		&=
		\begin{cases}
			S_{\text{boundary}(1)}+\L.\frac{\ l\,\mathrm{arctanh}\left(\frac{\ \sqrt{A^2l^2-1}\sqrt{1-x^2}}{\ x}\right)}{\ 2G_{3}}\R|^{x=x_{1}}_{x=x_{2}} & i=1, \; \rm for \; I_{\rm b1}, \\
			\\
			S_{\text{boundary}(2)}+\L.\frac{\ l\,\mathrm{arctanh}\left(\frac{\ \sqrt{A^2l^2+1}\sqrt{x^2-1}}{\ x}\right)}{\ 2 G_{3}}\R|^{x=x_{2}}_{x=x_{1}} & i=2, \; \rm for \; II_{\rm b1}\; and\; II_{\rm b2}, \\
			\\
			S_{\text{boundary}(3)}+\L.\frac{\ l\,\mathrm{arctanh}\left(\frac{\ x}{\ \sqrt{1-A^2l^2}\sqrt{1+x^2}}\right)}{\ 2G_{3}}\R|^{x=x_{2}}_{x=x_{1}} & i=3, \; \rm for \; III_{\rm b}.
		\end{cases}\\
		&=\alpha_{i}(\tilde{S}_{i}+\tilde{S}_{\text{boundary}_{i}}).
	\end{align}
	Thus we recover the results in \cite{previouswork} in the new theory. Compared with the thermodynamic relation in \cite{tian2024aspects}, in the new theory, we have:
	\begin{equation}
		2\tilde{\beta}_{i}M_{i}=\tilde{S}_{i}.
	\end{equation}
	\subsection{Holographic RG Flow}
	\quad In CFT, the central charge describes the degrees of freedom (DOF). Under RG flow, the DOF is monotonically decreasing, which is formally named as $c$-theorem:
	\begin{equation}\label{the derivative of g-funtion}
		c(\lambda)\geq 0,\; \frac{\mathrm{d}c(\lambda)}{\mathrm{d}\lambda}\leq 0.
	\end{equation}
	It points out the direction of RG flow. In AdS/BCFT, there is a similar version of $c$-theorem called $g$-theorem, and the boundary entropy can be used to give a candidate of $g$-function\cite{takayanagi2011holographic}. Given a boundary state $|B_{\alpha} \rangle$ with a boundary condition $g$ and a vacuum state $|0\rangle$, the boundary entropy is:
	\begin{equation}
		S^{\alpha}_{\rm boundary}=\log\, g_{\alpha}\, ,\quad g_{\alpha}=\langle 0|B_{\alpha}\rangle.
	\end{equation}
	In Section \ref{thermocmetric}, the boundary entropies of three classes in C-metric are formally represented as:
	\begin{equation}\label{3n boundary entropy}
		S_{\rm boundary}=\frac{l}{2G_{3}}\L.\mathrm{arctanh}\left(AlQ(x)\right)\R|^{x=x_{1}}_{x=x_{2}},
	\end{equation}
	accordingly we can define a $g$-function as
	\begin{equation}\label{loggfunc}
		\log g=f(z)=\frac{l}{2G_{3}}\L.\mathrm{arctanh}\left(AlQ(x(z))\right)\R|^{x=x_{1}}_{x=x_{2}}
	\end{equation}
	and test the $g$-theorem. Here $z$ is the radial coordinate in the FG expansion and $z=0$ corresponds to the boundary. Then similar to \eqref{the derivative of g-funtion} the derivative is
	\begin{equation}
		\frac{\mathrm{d}g(z)}{\mathrm{d}z}=g(z)\frac{\mathrm{d}f(z)}{\mathrm{d}z},
	\end{equation}
	thus we only focus on the monotonicity of $f(z)$. According to \eqref{loggfunc}, the derivative of $f(z)$ is:
	\begin{equation}\label{derivative of boundary entropy}
		\frac{\mathrm{d}f(z)}{\mathrm{d}z}=\L.\frac{l}{2G_{3}}\frac{AlQ'(x(z))x'(z)}{1-A^2l^2Q(x(z))}\R|^{x=x_{1}}_{x=x_{2}}.
	\end{equation}
	The key point is to find out the expression of  the function $x(z)$, however it's difficult in general cases. Then we'll consider this question and talk about the FG expansion in small acceleration $A$. As we can see in \eqref{3dCmetric2}, when $A$ is zero, $x$ and $y$ are decoupled, thus it's nature to expect these two directions are decoupled in the FG expansion. Here we also adopt ADM gauge and set Weyl factor $\omega(\xi)=1$, and the FG expansion in generic gauges will be discussed in the appendix \ref{appendixs0a}.
	The following are the results of the three classes.
	\begin{itemize}
		\item \textbf{Class I}: In the expansion of $x(z)$ \eqref{Transformation} we only keep the terms in the first order of the acceleration $A$, from the first several terms in the expansion we can find out the  expression of $x(z)$ which explicitly is
		\begin{equation}
			\begin{aligned}\label{x expansion of Class I}
				x=&\xi-Al(1-\xi^2)\(\frac{z}{l}\)-Al(1-\xi^2)\frac{1}{4}\(\frac{z}{l}\)^3-Al(1-\xi^2)\frac{1}{16}\(\frac{z}{l}\)^{5}-Al(1-\xi^2)\frac{1}{64}\(\frac{z}{l}\)^{7}\\
				&-Al(1-\xi^2)\frac{1}{256}\(\frac{z}{l}\)^{9}-...\\
				=&\xi-Az(1-\xi^2)\frac{1}{1-\frac{z^2}{4l^2}}, \,  (z^2<4l^2).
			\end{aligned}
		\end{equation}
		We expand the result of \eqref{FG in Class I} over small acceleration $A$ as:
		\begin{equation}
			\mathrm{d}s^2=-\frac{\ \left(z^2+4l^2\right)^2}{\ 16  l^2 z^2}\mathrm{d}\tau^2+\frac{\ l^2}{\ z^2}\mathrm{d}z^2+\frac{\ l^2\left(z^2-4l^2\right)^2}{\ 16z^2(1-\xi^2)}\mathrm{d}\xi^2+g(\mathcal{O}(A^2))_{ij}\mathrm{d}x^{i}\mathrm{d}x^{j}.
		\end{equation}
		The $z$ direction is truncated by $g_{\xi\xi}$ at $z^2=4l^2$. Considering $z=0$ (conformal boundary) is the inner point of the manifold, we adopt the region $z^2<4l^2$ of $z$, which ensures the convergence of expansion in \eqref{x expansion of Class I}. At this time, Class I is a particle with horizon $y_{\rm h_1}$ vanished and $x_{1},\, x_{2}\in [-1,1]$, but we can still discuss its boundary entropy. Then we substitute \eqref{x expansion of Class I} into \eqref{3n boundary entropy}, \eqref{derivative of boundary entropy} with the explicit expressions of $Q(x(z))$ in Section \ref{thermocmetric}, then we can define the $g$-function as
		\begin{equation}
			\log g(z)=f(z)=\frac{l}{2G_{3}}\(\mathrm{arctanh}\(Al\sqrt{1-x_{1}(z)^2}\)-\mathrm{arctanh}\(Al\sqrt{1-x_{2}(z)^2}\)\),
		\end{equation}
		and the corresponding derivative of $f(z)$ is: 
		\begin{align}
			\frac{\mathrm{d}f(z)}{\mathrm{d}z}&=\frac{2A^2l^4(4l^2+z^2)}{G_{3}(z^2-4l^2)^2}\(\xi_{1}\sqrt{1-\xi_{1}^2}-\xi_{2}\sqrt{1-\xi_{2}^2}\),\\
			&\approx\frac{2A^2l^4(4l^2+z^2)}{G_{3}(z^2-4l^2)^2}\(x_{1}\sqrt{1-x_{1}^2}-x_{2}\sqrt{1-x_{2}^2}\),
		\end{align}
		where we neglect the higher-order terms of $A$. The two functions $y=\sqrt{1-x^2}$ and $y=x\sqrt{1-x^2}$ are shown in Fig.\ref{Two function of C1}.
		\begin{figure}[h]
			\centering
			\includegraphics[width=0.40\linewidth]{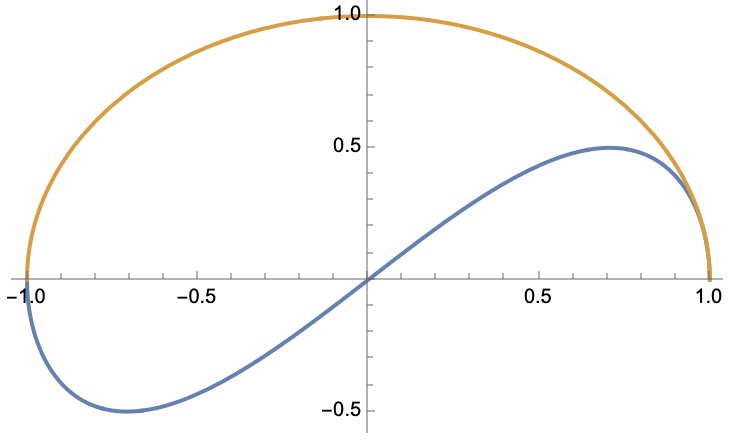}
			\caption{The picture of two functions. The orange curve is $y=\sqrt{1-x^2}$ and the blue curve is $y=x\sqrt{1-x^2}$. The extremal value point of $y=x\sqrt{1-x^2}$ is $x=\pm \frac{\sqrt{2}}{2}$.}
			\label{Two function of C1}
		\end{figure}
		According to the signs of $f(z)$ and $\frac{\mathrm{d}f(z)}{\mathrm{d}z}$, there exist four cases denoted as $\rm I_1$, $\rm I_2$, $\rm I_3$ and $\rm I_4$ which are demonstrated in Tab. \ref{tab:entropy_ranges}.
		\begin{table}[htbp]
			\centering
			\caption{Ranges of $x_{1},x_{2}$ for boundary entropy $f(z)$ and its derivative $f'(z)$ in Class I}
			\label{tab:entropy_ranges}
			\begin{tabular}{ccc>{\raggedright\arraybackslash}p{7.5cm}}
				\toprule
				Name & \textbf{$f(z)$} & \textbf{$f'(z)$} & \textbf{Valid Range $(x_1, x_2)$} \\
				\midrule
				$\rm I_1$ & $f(z) > 0$ & $f'(z) > 0$ & 
				$\begin{aligned}
					&\left(0 < x_1 \leq \tfrac{1}{\sqrt{2}} \land \sqrt{1 - x_1^2} < x_2 \leq 1\right) \\
					&\lor\ \left(\tfrac{1}{\sqrt{2}} < x_1 < 1 \land x_1 < x_2 \leq 1\right)
				\end{aligned}$ \\
				\cmidrule(lr){1-4}
				
				$\rm I_2$ & $f(z) > 0$ & $f'(z) < 0$ & 
				$\begin{aligned}
					&\left(-1 < x_1 < 0 \land -x_1 < x_2 \leq 1\right) \\
					&\lor\ \left(0 \leq x_1 < \tfrac{1}{\sqrt{2}} \land x_1 < x_2 < \sqrt{1 - x_1^2}\right)
				\end{aligned}$ \\
				\cmidrule(lr){1-4}
				
				$\rm I_3$ & $f(z) < 0$ & $f'(z) > 0$ & 
				$-1 \leq x_1 < -\tfrac{1}{\sqrt{2}} \land x_1 < x_2 < -\sqrt{1 - x_1^2}$ \\
				\cmidrule(lr){1-4}
				
				$\rm I_4$ & $f(z) < 0$ & $f'(z) < 0$ & 
				$\begin{aligned}
					&\left(-1 \leq x_1 < -\tfrac{1}{\sqrt{2}} \land -\sqrt{1 - x_1^2} < x_2 < -x_1\right) \\
					&\lor\ \left(-\tfrac{1}{\sqrt{2}} \leq x_1 < 0 \land x_1 < x_2 < -x_1\right)
				\end{aligned}$ \\
				\bottomrule
			\end{tabular}
		\end{table}
		
		\item \textbf{Class II}: In parallel, we keep the expansion of $x(z)$ to the first order of the acceleration $A$, the result is:
		\begin{equation}
			\begin{aligned}\label{x expansion of Class II}
				x=&\xi-Al(\xi^2-1)\(\frac{z}{l}\)+Al(\xi^2-1)\frac{1}{4}\(\frac{z}{l}\)^3-Al(\xi^2-1)\frac{1}{16}\(\frac{z}{l}\)^{5}+Al(\xi^2-1)\frac{1}{64}\(\frac{z}{l}\)^{7}\\
				&-Al(\xi^2-1)\frac{1}{256}\(\frac{z}{l}\)^{9}-...\\
				=&\xi-Az(\xi^2-1)\frac{1}{1+\frac{z^2}{4l^2}}, \,  (z^2<4l^2).
			\end{aligned}
		\end{equation}
		Then we expand the result of \eqref{FG in Class II} over small acceleration as:
		\begin{equation}
			\mathrm{d}s^2=-\frac{\ (z^2-4l^2)^2}{16  l^2 z^2}\mathrm{d}\tau^2 +\frac{\ l^2}{\ z^2}\mathrm{d}z^2+\frac{\ (z^2+4l^2)^2}{\ 16z^2(\xi^2-1)}\mathrm{d}\xi^2+g(\mathcal{O}(A^2))_{ij}\mathrm{d}x^{i}\mathrm{d}x^{j}.
		\end{equation} 
		The $z$ direction is truncated by $g_{\tau\tau}$ at $z^2=4l^2$. Considering $z=0$ (conformal boundary) is the inner point of the manifold, we adopt the region $z^2<4l^2$ of $z$, which ensures the convergence of the expansion in \eqref{x expansion of Class II}. Next we substitute \eqref{x expansion of Class II} into \eqref{derivative of boundary entropy} and $g$-function as 
		\begin{equation}
			\log g(z)=f(z)=\frac{l}{2G_{3}}\(\mathrm{arctanh}\(Al\sqrt{x_{1}^2-1}\)-\mathrm{arctanh}\(Al\sqrt{x_{2}^2-1}\)\).
		\end{equation}
		The derivative of $f(z)$ is:
		\begin{equation}
			f'(z)\approx-\frac{2A^2l^4(4l^2-z^2)}{G_{3}(z^2+4l^2)^2}\(x_{1}\sqrt{x_{1}^2-1}-x_{2}\sqrt{x_{2}^2-1}\),
		\end{equation}
		where we also neglect the higher-order terms of $A$. The functions of $y=\sqrt{x^2-1}$ and $y=-x\sqrt{x^2-1}$ are shown in Fig. \ref{Two function of C2}.
		\begin{figure}[h]
			\centering
			\includegraphics[width=0.40\linewidth]{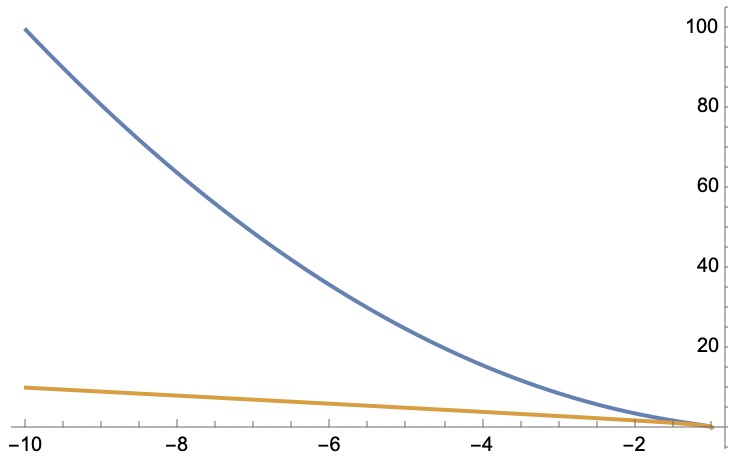}
			\centering
			\includegraphics[width=0.40\linewidth]{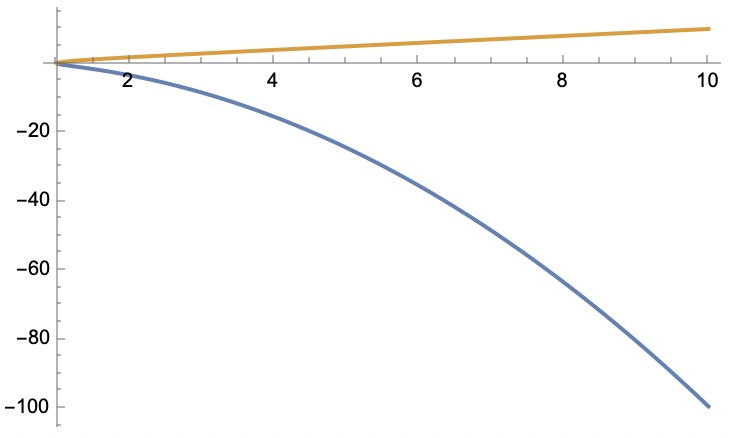}
			\caption{\text{Left}: The pictures of two functions with $x<-1$. \text{Right}: The pictures of two functions with $x>1$.  The orange curve is $y=\sqrt{x^2-1}$ and the blue curve is $y=-x\sqrt{x^2-1}$.}
			\label{Two function of C2}
		\end{figure}
		According to the signs of $f(z)$ and $\frac{\mathrm{d}f(z)}{\mathrm{d}z}$, there exist two cases and we denote them as $\rm II_1$, $\rm II_2$ which are demonstrated in Fig. \ref{tab:entropy_ranges2}.
		\begin{table}[htbp]
			\centering
			\caption{Ranges $x_{1},x_{2}$ for boundary entropy $f(z)$ and its derivative $f'(z)$ in Class II}
			\label{tab:entropy_ranges2}
			\begin{tabular}{ccc>{\raggedright\arraybackslash}p{7.5cm}}
				\toprule
				Name & \textbf{$f(z)$} & \textbf{$f'(z)$} & \textbf{Valid Range $(x_1, x_2)$} \\
				\midrule
				$\rm II_1$ & $f(z) > 0$ & $f'(z) > 0$ & 
				$-y_{\rm h_2}<x_1<x_2\leq -1$ \\
				\cmidrule(lr){1-4}
				
				$\rm II_2$ & $f(z) < 0$ & $f'(z) > 0$ & 
				$1\leq x_1<x_2<y_{\rm h_2}$ \\
				\bottomrule
			\end{tabular}
		\end{table}
		
		\item \textbf{Class III}: Similarly, we keep the expansion of $x(z)$ to the first order of the acceleration $A$, the result is:
		\begin{equation}
			\begin{aligned}\label{x expansion of Class III}
				x=&\xi-Al(\xi^2+1)\(\frac{z}{l}\)+Al(\xi^2+1)\frac{1}{4}\(\frac{z}{l}\)^3-Al(\xi^2+1)\frac{1}{16}\(\frac{z}{l}\)^{5}+Al(\xi^2+1)\frac{1}{64}\(\frac{z}{l}\)^{7}\\
				&-Al(\xi^2+1)\frac{1}{256}\(\frac{z}{l}\)^{9}-...\\
				=&\xi-Az(\xi^2+1)\frac{1}{1+\frac{z^2}{4l^2}}, \,  (z^2<4l^2).
			\end{aligned}
		\end{equation}
		Then we expand the result of \eqref{FG in Class III} over small acceleration as:
		\begin{equation}
			\mathrm{d}s^2=-\frac{\ (z^2-4l^2)^2}{16  l^2 z^2}\mathrm{d}\tau^2 +\frac{\ l^2}{\ z^2}\mathrm{d}z^2+\frac{\ (z^2+4l^2)^2}{\ 16z^2(\xi^2+1)}\mathrm{d}\xi^2+g(\mathcal{O}(A^2))_{ij}\mathrm{d}x^{i}\mathrm{d}x^{j}.
		\end{equation} 
		The $z$ direction is truncated by $g_{\tau\tau}$ at $z^2=4l^2$. Considering $z=0$ (conformal boundary) is the inner point of the manifold, we adopt the region $z^2<4l^2$ of $z$, which ensures the convergence of the expansion in \eqref{x expansion of Class II}. Next we substitute \eqref{x expansion of Class II} into \eqref{derivative of boundary entropy} and define the $g$-function as 
		\begin{equation}
			\log g(z)=f(z)=\frac{l}{2G_{3}}\(\mathrm{arctanh}\(Al\sqrt{x_{1}^2+1}\)-\mathrm{arctanh}\(Al\sqrt{x_{2}^2+1}\)\).
		\end{equation}
		The derivative of $f(z)$ is
		\begin{equation}
			f'(z)\approx-\frac{2A^2l^4(4l^2-z^2)}{G_{3}(z^2+4l^2)^2}\(x_{1}\sqrt{x_{1}^2+1}-x_{2}\sqrt{x_{2}^2+1}\),
		\end{equation}
		where we also neglect the higher-order terms of $A$.
		The functions of $y=\sqrt{x^2+1}$ and $y=-x\sqrt{x^2+1}$ is shown in Fig. \ref{Two function of C3}.
		\begin{figure}[h]
			\centering
			\includegraphics[width=0.45\linewidth]{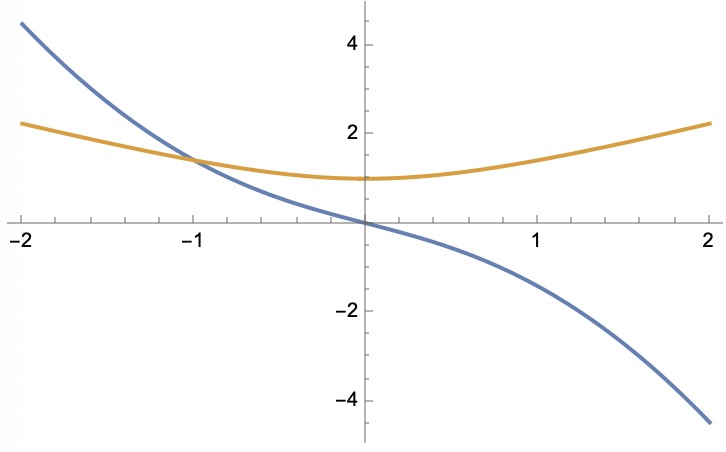}
			\caption{The pictures of two functions.  The orange curve is $y=\sqrt{x^2+1}$ and the blue curve is $y=-x\sqrt{x^2+1}$. The intersection point is $x=-1$.}
			\label{Two function of C3}
		\end{figure}
		According to the signs of $f(z)$ and $\frac{\mathrm{d}f(z)}{\mathrm{d}z}$, there exist two cases and we denote them as $\rm III_1$, $\rm III_2$ which are demonstrated in Fig. \ref{tab:entropy_ranges3}.
		\begin{table}[htbp]
			\centering
			\caption{Ranges of $x_{1},x_{2}$ for boundary entropy $f(z)$ and its derivative $f'(z)$ in Class III}
			\label{tab:entropy_ranges3}
			\begin{tabular}{ccc>{\raggedright\arraybackslash}p{7.5cm}}
				\toprule
				Name & \textbf{$f(z)$} & \textbf{$f'(z)$} & \textbf{Valid Range $(x_1, x_2)$} \\
				\midrule
				$\rm III_1$ & $f(z) > 0$ & $f'(z) > 0$ & 
				$-y_{\rm h_3}<x_1<0 \land x_1<x_2< -x_1$ \\
				\cmidrule(lr){1-4}
				
				$\rm III_2$ & $f(z) < 0$ & $f'(z) > 0$ & 
				$\(-y_{\rm h_3}<x_1\leq 0 \land -x_1<x_2<y_{\rm h_3} \)\lor \(0<x_1 \land x_1<x_2<y_{\rm h_3}\)$\\
				\bottomrule
			\end{tabular}
		\end{table}
	\end{itemize}
	
	As we can see above, in $\rm I_3$, $\rm I_4$, $\rm II_2$ and $\rm III_2$, $f(z)<0$, thus these cases disobey the null energy condition. In $\rm I_1$, $\rm I_3$, $\rm II_1$, $\rm II_2$, $\rm III_1$ and $\rm III_3$, from UV to IR $f'(z)>0$, it means that the boundary entropy increases along the direction of RG flow. As stated in \cite{Affleck:1991tk}, the DOF degenerates from UV to IR, thus the derivative of $f(z)$ is expected to be less than 0. In all cases, only in $\rm I_2$ the null energy condition and $f'(z)<0$ are both satisfied, in which the $g$-theorem can be derived. There may exist exceptions that the boundary entropy increases under bulk RG flow \cite{Green:2007wr}. The negative boundary entropy and the increasing boundary entropy under RG flow in C-metric need more physical interpretations, we can investigate these questions in the future.   
	\section{Conclusion}\label{conclusion}
	In this work, we re-examine the computation of counting the microstates in BTZ black holes and extend it to the case of 3D C-metric. As a setup we discuss the case of a sector that is cut from a BTZ black hole, and introduce a new theory by lifting this sector to the covering space and redefining the mass parameter and Newton's constant, after which the angular direction is restored to $[0,2\pi]$ in this new theory. Accordingly, we modify the Cardy formula, by which the microscopic derivation of the black hole entropy satisfies the area law. Furthermore, we apply this method to express the entropy of the accelerating BTZ black hole by Virasoro algebra in the new theory.  Finally in small acceleration, we discuss the holographic RG flow in the three solutions of C-metric, and find that only $\rm I_2$ is physically sensible currently, in which $g$-theorem can be derived.
	
	\section*{Acknowledgement}
	We would like to thank Jia Tian, Wenbin Pan, Jiayin Shen and Liangyu Chen for constructive discussion and advice.  SX and LL are
	supported by the  NSFC NO. 12473038.
	
	\appendix
	\section{The Degeneration of the  FG Expansion with A=0} \label{appendixs0a}
	In this part, we promote the FG expansion of C-metric to cases with $A=0$ in generic gauges.
	
	For Class I, when $A=0$, the metric becomes:
	\begin{equation}\label{Class I with A=0}
		\mathrm{d}s^2=\frac{1}{y^2}\left[-(\frac{1}{l^2}+y^2)\mathrm{d}\tau^2+\frac{\mathrm{d}y^2}{\frac{1}{l^2}+y^2}+\frac{\mathrm{d}x^2}{1-x^2} \right],
	\end{equation}
	and the transformation of an infinite-ordered polynomial is:
	\begin{equation}
		\begin{aligned}
			y&=\sum^{\infty}_{m=0}\tilde{F}_{2m+1}(\xi)\left(\frac{z}{l}\right)^{2m+1}, \quad x=\xi+ \sum^{\infty}_{m=1}\tilde{G}_{2m}(\xi)\left(\frac{z}{l}\right)^{2m},\\
			\tilde{F}_{1}&=\frac{1}{l\Omega(\xi)}, \quad \tilde{F}_{3}=\frac{\Omega(\xi)^2-(1-\xi^2)\Omega'(\xi)^2}{4l\Omega(\xi)^5},\, ...\\
			\tilde{G}_{2}&=\frac{(1-\xi^2)\Omega'(\xi)}{2\Omega(\xi)^3}, \quad \tilde{G}_{4}=\frac{(1-\xi^2)(\Omega(\xi)^2\Omega'(\xi)-\xi\Omega(\xi)\Omega'(\xi)^2-(1-\xi^2)\Omega'(\xi)^3)}{\Omega(\xi)^7},\, ....
		\end{aligned}
	\end{equation}
	Following \cite{previouswork}, we make a substitution 
	\begin{align}
		\Xi=l\,\mathrm{arccos}(\xi),\quad \tilde{\Omega}(\Xi)=\Omega(\xi),
	\end{align}
	then line element \eqref{Class I with A=0} becomes:
	\begin{equation}
		\begin{aligned}\label{FG of class I}
			\mathrm{d}s^2=    &-\frac{(z^2\tilde{\Omega}(\Xi)^2+4l^2\tilde{\Omega}(\Xi)^4+z^2l^2\tilde{\Omega}'(\Xi)^2)^2}{16l^2z^2\tilde{\Omega}(\Xi)^6}\mathrm{d}\tau^2+\frac{l^2}{z^2}\mathrm{d}z^2\\
			&\quad+\frac{(-\tilde{\Omega}(\Xi)^2z^2+4l^2\tilde{\Omega}(\Xi)^4-3l^2\tilde{\Omega}'(\Xi)^2z^2+2l^2\tilde{\Omega}(\Xi)\tilde{\Omega}''(\Xi)z^2)^2}{16l^2\tilde{\Omega}(\Xi)^6z^2}\mathrm{d}\Xi^2.
		\end{aligned}
	\end{equation}
	
	For Class II, when $A=0$, the metric becomes:
	\begin{equation}\label{Class II with A=0}
		\mathrm{d}s^2=\frac{1}{y^2}\left[-(\frac{1}{l^2}-y^2)\mathrm{d}\tau^2+\frac{\mathrm{d}y^2}{\frac{1}{l^2}-y^2}+\frac{\mathrm{d}x^2}{1-x^2} \right],
	\end{equation}
	and the transformation of an infinite-ordered polynomial is:
	\begin{equation}
		\begin{aligned}
			y&=\sum^{\infty}_{m=0}\tilde{F}_{2m+1}(\xi)\left(\frac{z}{l}\right)^{2m+1}, \quad x=\xi+ \sum^{\infty}_{m=1}\tilde{G}_{2m}(\xi)\left(\frac{z}{l}\right)^{2m},\\
			\tilde{F}_{1}&=\frac{1}{l\Omega(\xi)}, \quad \tilde{F}_{3}=-\frac{\Omega(\xi)^2+(\xi^2-1)\Omega'(\xi)^2}{4l\Omega(\xi)^5},\, ...\\
			\tilde{G}_{2}&=\frac{(\xi^2-1)\Omega'(\xi)}{2\Omega(\xi)^3}, \quad \tilde{G}_{4}=\frac{(\xi^2-1)(-\Omega(\xi)^2\Omega'(\xi)+\xi\Omega(\xi)\Omega'(\xi)^2-(\xi^2-1)\Omega'(\xi)^3)}{\Omega(\xi)^7},\, ....
		\end{aligned}
	\end{equation}
	Similarly, we make a substitution
	\begin{align}
		\Xi=l\,\mathrm{arccosh}(|\xi|),\quad \tilde{\Omega}(\Xi)=\Omega(\xi),
	\end{align}
	then the line element \eqref{Class II with A=0} becomes:
	\begin{equation}
		\begin{aligned}\label{FG of class II}
			\mathrm{d}s^2=&-\frac{(z^2\tilde{\Omega}(\Xi)^2-4l^2\tilde{\Omega}(\Xi)^4-z^2l^2\tilde{\Omega}'(\Xi)^2)^2}{16l^2z^2\tilde{\Omega}(\Xi)^6}\mathrm{d}\tau^2+\frac{l^2}{z^2}\mathrm{d}z^2\\
			&\quad+\frac{(\tilde{\Omega}(\Xi)^2z^2+4l^2\tilde{\Omega}(\Xi)^4-3l^2\tilde{\Omega}'(\Xi)^2z^2+2l^2\tilde{\Omega}(\Xi)\tilde{\Omega}''(\Xi)z^2)^2}{16l^2\tilde{\Omega}(\Xi)^6z^2}\mathrm{d}\Xi^2.
		\end{aligned}
	\end{equation}
	
	For Class III, when $A=0$, the metric also becomes:
	\begin{equation}\label{Class II with A=0}
		\mathrm{d}s^2=\frac{1}{y^2}\left[-(\frac{1}{l^2}-y^2)\mathrm{d}\tau^2+\frac{\mathrm{d}y^2}{\frac{1}{l^2}-y^2}+\frac{\mathrm{d}x^2}{1-x^2} \right],
	\end{equation}
	and the transformation of an infinite-ordered polynomial is:
	\begin{equation}
		\begin{aligned}
			y&=\sum^{\infty}_{m=0}\tilde{F}_{2m+1}(\xi)\left(\frac{z}{l}\right)^{2m+1}, \quad x=\xi+ \sum^{\infty}_{m=1}\tilde{G}_{2m}(\xi)\left(\frac{z}{l}\right)^{2m},\\
			\tilde{F}_{1}&=\frac{1}{l\Omega(\xi)}, \quad \tilde{F}_{3}=-\frac{\Omega(\xi)^2+(\xi^2+1)\Omega'(\xi)^2}{4l\Omega(\xi)^5},\, ...\\
			\tilde{G}_{2}&=\frac{(\xi^2+1)\Omega'(\xi)}{2\Omega(\xi)^3}, \quad \tilde{G}_{4}=-\frac{(\xi^2+1)(\Omega(\xi)^2\Omega'(\xi)-\xi\Omega(\xi)\Omega'(\xi)^2+(\xi^2+1)\Omega'(\xi)^3)}{\Omega(\xi)^7},\, ....
		\end{aligned}
	\end{equation}
	Similarly, we make a substitution 
	\begin{align}
		\Xi=l\,\mathrm{arcsinh}(\xi),\quad \tilde{\Omega}(\Xi)=\Omega(\xi),
	\end{align}
	then the line element \eqref{Class II with A=0} becomes:
	\begin{equation}
		\begin{aligned}\label{FG of class III}
			\mathrm{d}s^2&=-\frac{(z^2\Omega(\Xi)^2-4l^2\Omega(\Xi)^4-z^2l^2\Omega'(\Xi)^2)^2}{16l^2z^2\Omega(\Xi)^6}\mathrm{d}\tau^2+\frac{l^2}{z^2}\mathrm{d}z^2\\
			&\quad+\frac{(\Omega(\Xi)^2z^2+4l^2\Omega(\Xi)^4-3l^2\Omega'(\Xi)^2z^2+2l^2\Omega(\Xi)\Omega''(\Xi)z^2)^2}{16l^2\Omega(\Xi)^6z^2}\mathrm{d}\Xi^2.
		\end{aligned}
	\end{equation}
	
	As we can see above, only when $\Omega(\xi)=\text{constant}$, $x$ and $y$ decouple in the transformation. In AdS$_{3}$ without acceleration, we can also introduce Weyl gauge in FG expansion.  \eqref{FG of class I} gives the FG expansion of thermal AdS (or particle), while \eqref{FG of class II} and \eqref{FG of class III} give the FG expansion of BTZ black hole\footnote{The FG expansion of thermal AdS and BTZ black hole can also be obtained from the general metric of AdS$_3$ by making a transformation and setting a special values of $b$ with no rotation in \cite{Skenderis:1999nb}.}. Near the boundary, \eqref{FG of class I}, \eqref{FG of class II} and \eqref{FG of class III} have similar form:
	\begin{equation}
		\mathrm{d}s^2=\frac{l^2}{z^2}\mathrm{d}z^2+\frac{g_{ij}(z,\Xi)}{z^2}\mathrm{d}x^{i}\mathrm{d}x^{j}=\frac{l^2}{z^2}\mathrm{d}z^2+\frac{l^2\tilde{\Omega}(\Xi)^2(-\mathrm{d}\tau^2+\mathrm{d}\Xi^2)}{z^2}+\frac{g_{ij}(\mathcal{O}(z^{2}))}{z^2}\mathrm{d}x^{i}\mathrm{d}x^{j}+....
	\end{equation}
	We set a cut-off as $z=\epsilon\ll 1$ and calculate the 2D Ricci scalar and Ricci tensor of $g_{ij}$, then we can get:
	\begin{equation}
		\begin{aligned}
			R(g_{ij})&=\frac{2(\tilde{\Omega}'(\Xi)^2-\tilde{\Omega}(\Xi)\tilde{\Omega}''(\Xi))}{l^2\tilde{\Omega}(\Xi)^4},\\
			R_{ij}(g_{ij})&=\frac{(\tilde{\Omega}'(\Xi)^2-\tilde{\Omega}(\Xi)\tilde{\Omega}''(\Xi))}{l^2\tilde{\Omega}(\Xi)^4}h_{ij}.
		\end{aligned}
	\end{equation}
	When $\tilde{\Omega}(\Xi)=e^{\alpha+\mu\Xi}\; (\alpha,\mu\in\mathbb{R})$, the Ricci curvature vanishes, and the background of the boundary CFT is flat, otherwise it is a curved spacetime. More discussions can be found in \cite{Li:2025rzl}. The stress tensors of the thermal AdS (or particle\footnote{Compared with vacuum AdS, the particle has angular defect.}) and BTZ black hole are:
		\begin{equation}
			\begin{aligned}\label{Stress tensor}
				T^{\tau}_{\tau(\rm particle)}&=-\frac{\tilde{\Omega}(\Xi)^2+3l^2\tilde{\Omega}'(\Xi)^2-2l^2\tilde{\Omega}(\Xi)\tilde{\Omega}''(\Xi)}{16G_{3}\pi l\tilde{\Omega}(\Xi)^4},\\
				T^{\Xi}_{\Xi(\rm particle)}&=\frac{\tilde{\Omega}(\Xi)^2+l^2(\tilde{\Omega}'(\Xi))^2}{16G_{3}\pi l\tilde{\Omega}(\Xi)^4},\\
				T^{\tau}_{\tau(\rm BTZ)}&=\frac{\tilde{\Omega}(\Xi)^2-3l^2\tilde{\Omega}'(\Xi)^2+2l^2\tilde{\Omega}(\Xi)\tilde{\Omega}''(\Xi)}{16G_{3}\pi l\tilde{\Omega}(\Xi)^4},\\
				T^{\Xi}_{\Xi(\rm BTZ)}&=-\frac{\tilde{\Omega}(\Xi)^2-l^2(\tilde{\Omega}'(\Xi))^2}{16G_{3}\pi l\tilde{\Omega}(\Xi)^4}.
			\end{aligned}
		\end{equation}
		The trace of the tress tensor gives:
	\begin{equation}
		T^{i}_{i}=\frac{c}{24}R(g_{ij}).
	\end{equation}
	
	Following \cite{previouswork}, we set the boundary condition as follows:
	\begin{equation}
		\delta g={\begin{bmatrix}
				\mathcal{O}(1)&\mathcal{O}\(\frac{1}{z}\)&\mathcal{O}(1)\\
				\mathcal{O}\(\frac{1}{z}\)&0&\mathcal{O}\(\frac{1}{z}\)\\
				\mathcal{O}(1)&\mathcal{O}\(\frac{1}{z}\)&\mathcal{O}(1)\\
		\end{bmatrix}},
	\end{equation}
	The asymptotic killing vector is:
	\begin{equation}
		\begin{aligned}\label{B.7}
			\mathcal{X}^{(3)}=&\(\frac{T(t+\Xi)+M(t-\Xi)}{2}\)\partial_{t}+\(\frac{z\partial_{\Xi}(\Omega(\Xi)(T(t+\Xi)-M(t-\Xi)))}{2\Omega(\Xi)}\)\partial_{z}\\
			&\quad+\(\frac{T(t+\Xi)-M(t-\Xi)}{2}\)\partial_{\Xi},
		\end{aligned}
	\end{equation}
	$T(t+\Xi)$ and $M(t-\Xi)$ are arbitrary functions and correspond to two modes. Here we also introduce an angular defect by building the equivalent relation $\Xi\sim\Xi+2\pi\beta$, the boundary CFT is bounded, but is not coupled with the defect with no acceleration compared with C-metric. Considering the periodicity of $\Xi$, we introduce two Fourier modes $e^{-i n (t+\Xi)/\beta}$ and $e^{-i n (t-\Xi)/\beta}$,  where $\beta\in\mathbb{R}/\{0\}$. Accordingly, the asymptotic algebras of two modes are given as:
	\begin{align}
		\mathcal{X}^{(3)}_{(R)n}&=\frac{\beta e^{-\frac{i n (t+\Xi)}{\beta}}}{2}\partial_{t}+\frac{z\beta \partial_{\Xi}\(\Omega(\Xi)e^{-\frac{i n (t+\Xi)}{\beta}}\)}{2\Omega(\Xi)}\partial_{z}+\frac{\beta e^{-\frac{i n (t+\Xi)}{\beta}}}{2}\partial_{\Xi},\\
		\mathcal{X}^{(3)}_{(L)n}&=\frac{\beta e^{-\frac{i n (t-\Xi)}{\beta}}}{2}\partial_{t}-\frac{z\beta \partial_{\Xi}\(\Omega(\Xi)e^{-\frac{i n (t-\Xi)}{\beta}}\)}{2\Omega(\Xi)}\partial_{z}-\frac{\beta e^{-\frac{i n (t-\Xi)}{\beta}}}{2}\partial_{\Xi},
	\end{align}
	they meet the classical commutation relation:
		\begin{align}
			[ \mathcal{X}^{(3)}_{(R)n},\mathcal{X}^{(3)}_{(R)m}]&=i(n-m)\mathcal{X}^{(3)}_{(R)n+m},\\
			[ \mathcal{X}^{(3)}_{(L)n},\mathcal{X}^{(3)}_{(L)m}]&=i(n-m)\mathcal{X}^{(3)}_{(L)n+m},\\
			[ \mathcal{X}^{(3)}_{(R)n},\mathcal{X}^{(3)}_{(L)m}]&=0.
		\end{align}
	Here we define a set $\mathbb{W}$ consisting of all generators as:
		\begin{equation}
			\mathbb{W}:=\{\lambda_{m}\mathcal{X}^{(3)}_{(L)m}+\lambda_{n}\mathcal{X}^{(3)}_{(R)n}|m,n\in\mathbb{Z},\lambda_{m}\in\mathbb{C},\lambda_{n}\in\mathbb{C}\}.
	\end{equation}
	Given a 2+1 decomposition as
	\begin{equation}\label{2+1 decomposition}
		\mathrm{d}s^2=G_{\mu\nu}\mathrm{d}x^{\mu}\mathrm{d}x^{\nu}=-N^2\mathrm{d}t^2+h_{ij}(N^{i}\mathrm{d}t+\mathrm{d}x^{i})(N^{j}\mathrm{d}t+\mathrm{d}x^{j})\, ,
	\end{equation}
	where $N$ is the lapse function and $N^{i}$ is the shift function. The conjugate variables are
	\begin{equation}
		K_{ij}=\frac{\dot{h}_{ij}-\nabla_{i}N_{j}-\nabla_{j}N_{i}}{2N}\,, \qquad 
		\Pi^{ij}=\sqrt{h}(K^{ij}-Kh^{ij})\,,
	\end{equation}
	where $h$ represents the spatial part of the metric  and the dot represents the derivative with respect to time.
	
 We define a set that consists of all the directions of the deformation that generates the Hamiltonian as:
		\begin{equation}
			\tilde{\mathbb{W}}:=\{P^{\mu}_{\alpha}\mathcal{X}^{(3)\alpha}=\(N\mathcal{X}^{(3)t},\; \mathcal{X}^{(3)z},\;\mathcal{X}^{(3)\Xi}\)|\mathcal{X}^{(3)}\in\mathbb{W}\}.
		\end{equation}
		Following \cite{Brown:1986nw}, the extended central term is given as\footnote{The detailed calculations can be seen in the prepared paper \cite{previouswork}.}:
		\begin{equation}
			\begin{aligned}\label{extended term}
				&K(\mathcal{X},\mathcal{Y})\\
				=&-\frac{1}{16\pi G_{3}}\lim\limits_{z \to 0}\oint \mathrm{d}x_{l}\left[G^{ijkl}(h)(N\mathcal{X}^{(3)t}\nabla_{k}(\tilde{h}_{ij}-h_{ij})-\nabla_{k}(N\mathcal{X}^{(3)t})(\tilde{h}_{ij}-h_{ij}))+2\mathcal{X}^{(3)m}\Pi_{m}^{l}(\tilde{h})\right],
			\end{aligned}
		\end{equation}
		where $G^{ijkl}(h)=\sqrt{h}(h^{ik}h^{jl}-h^{ij}h^{kl})$ and $\tilde{h}_{ij}=h_{ij}+\mathcal{L}_{\mathcal{Y}}h_{ij}$. Near the boundary, we have
	\begin{align}
		N=\frac{l}{z}+\mathcal{O}(1), \quad N^{i}=0, \quad \Pi^{z}_{z}=\mathcal{O}(z), \quad \Pi^{z}_{\Xi}=\mathcal{O}(z^3).
	\end{align}
	After some computation, $\mathcal{X}^{m}\Pi_{m}^{l}(\tilde{h})\sim\mathcal{O}(z)$ can be neglected.

	For vacuum AdS$_{3}$ of \eqref{FG of class I}, ($\ref{extended term}$) gives:
	\begin{align}\label{Central term of I in generic gauge}
		K(\mathcal{X}_{(R/L)n},\mathcal{X}_{(R/L)m})        =&\delta_{m+n}\int^{\beta\pi}_{-\beta\pi}\(-\frac{iln^3}{16G_{3}\pi\beta}-\frac{in\beta(-2\tilde{\Omega}(\Xi)^2-l^2\tilde{\Omega}'(\Xi)^2+l^2\tilde{\Omega}(\Xi)\tilde{\Omega}''(\Xi))}{32G_{3}l\pi\tilde{\Omega}(\Xi)^2}\R.\nonumber\\
		&\L.-\frac{l\beta^2(2\tilde{\Omega}'(\Xi)^3-3\tilde{\Omega}(\Xi)\Omega'(\Xi)\tilde{\Omega}''(\Xi)+\tilde{\Omega}(\Xi)^2\tilde{\Omega}'''(\Xi))}{32G_{3}\pi\tilde{\Omega}(\Xi)^3}\)\mathrm{d}\Xi, 
	\end{align}
	in which $\delta_{m+n}=1$ with $m+n=0$ or $\delta_{m+n}=0$. As we can see above, the coefficient of $n^{3}$ gives the central charge $\frac{\ 3l}{\ 2G_{2}}$. To recover the Virasoro algebra, the third term of the right hand side in \eqref{Central term of I in generic gauge} is required to be 0, we have
	\begin{align}\label{Equation1}
		2\tilde{\Omega}'(\Xi)^3-3\tilde{\Omega}(\Xi)\tilde{\Omega}'(\Xi)\tilde{\Omega}''(\Xi)+\tilde{\Omega}(\Xi)^2\tilde{\Omega}'''(\Xi)&=0,\\
		\tilde{\Omega}(\Xi)(\ln(\tilde{\Omega}(\Xi)))'''&=0,\\
		(\ln(\tilde{\Omega}(\Xi)))'''&=0(\tilde{\Omega}(\Xi)\neq 0).
	\end{align}
	The solution $\tilde{\Omega}(\Xi)$ is:
	\begin{equation}\label{Solution1}
		\tilde{\Omega}(\Xi)= e^{\alpha+\mu\Xi+\gamma \Xi^2},
	\end{equation}
	with $\alpha$, $\mu$ and $\gamma$ the free constants. After applying \eqref{Solution1} to \eqref{Central term of I in generic gauge} we can get:
	\begin{equation}
		\begin{aligned}\label{Extended term in generic gauge of Class I}
			&K(\mathcal{X}_{(R/L)n},\mathcal{X}_{(R/L)m})\\
			=&-\frac{il}{8\pi G_{3}}\(n^3+ n\beta^2\(-\frac{1}{l^2}+\gamma\)\)\delta_{m+n}=-i\frac{c}{12}\(n^3+ n\beta^2\(-\frac{1}{l^2}+\gamma\)\)\delta_{m+n},
		\end{aligned}
	\end{equation}
	with $c=\frac{3l}{2G_{3}}$.
	
	For BTZ black hole of \eqref{FG of class II} or \eqref{FG of class III}, \eqref{extended term} gives:
	\begin{align}\label{Central term of II in generic gauge}
		K(\mathcal{X}_{(R/L)n},\mathcal{X}_{(R/L)m})    =&\delta_{m+n}\int^{\beta\pi}_{-\beta\pi}\(-\frac{iln^3}{16G_{3}\pi\beta}-\frac{in\beta(2\tilde{\Omega}(\Xi)^2-l^2\tilde{\Omega}'(\Xi)^2+l^2\tilde{\Omega}(\Xi)\tilde{\Omega}''(\Xi))}{32G_{3}l\pi\tilde{\Omega}(\Xi)^2}\R.\nonumber\\
		&\L.-\frac{l\beta^2(2\tilde{\Omega}'(\Xi)^3-3\tilde{\Omega}(\Xi)\tilde{\Omega}'(\Xi)\tilde{\Omega}''(\Xi)+\tilde{\Omega}(\Xi)^2\tilde{\Omega}'''(\Xi))}{32G_{3}\pi\tilde{\Omega}(\Xi)^3}\)\mathrm{d}\Xi,
	\end{align}
	in which $\delta_{m+n}=1$ with $m+n=0$ or $\delta_{m+n}=0$. Similarly, the central charge can be read off from the coefficient of $n^{3}$ as $\frac{\ 3l}{\ 2G_{3}}$. Then we require:
	\begin{equation}\label{Equation2}
		2\tilde{\Omega}'(\Xi)^3-3\tilde{\Omega}(\Xi)\tilde{\Omega}'(\Xi)\tilde{\Omega}''(\Xi)+\tilde{\Omega}(\Xi)^2\tilde{\Omega}'''(\Xi)=0,
	\end{equation}
	and the solution $\tilde{\Omega}(\Xi)$ is:
	\begin{equation}\label{Solution2}
		\tilde{\Omega}(\Xi)=e^{\alpha+\mu\Xi+\gamma \Xi^2},
	\end{equation}
	with $\alpha$, $\mu$ and $\gamma$ the free constants. After applying \eqref{Solution2} to \eqref{Central term of II in generic gauge}, we can also get:
	\begin{equation}
		\begin{aligned}\label{Extended term in generic gauge of Class II}
			&K(\mathcal{X}_{(R/L)n},\mathcal{X}_{(R/L)m})\\
			=&-\frac{il}{8\pi G_{3}}\(n^3+ n\beta^2\(\frac{1}{l^2}+\gamma\)\)\delta_{m+n}=-i\frac{c}{12}\(n^3+ n\beta^2\(\frac{1}{l^2}+\gamma\)\)\delta_{m+n},
		\end{aligned}
	\end{equation}
	with $c=\frac{3l}{2G_{3}}$. We can see above that coefficient $\gamma$ in $\tilde{\Omega}(\Xi)$ will change the zero-point energy.
	
	Furthermore, we employ $\tilde{\Omega}(\Xi)=e^{\alpha+\mu\Xi+\gamma\Xi^{2}}$ and calculate the holographic mass based on \eqref{holo-mass2}. The masses of the vaccum AdS and the BTZ black hole are:
			\begin{align}
				M_{\rm particle}&=-\oint_{2\pi\beta}  \mathrm{d}\Xi\frac{\tilde{\Omega}(\Xi)^2+3l^2\tilde{\Omega}'(\Xi)^2-2l^2\tilde{\Omega}(\Xi)\tilde{\Omega}''(\Xi)}{16G_{3}l\pi\tilde{\Omega}(\Xi)^2}\\
				&=-\frac{\beta(3+l^2(16\pi^2\beta^2\gamma^2+3\mu^2+12\gamma(\pi\beta\mu-1)))}{24G_{3}l},\\
				M_{\rm BTZ}&=\oint_{2\pi\beta} \mathrm{d}\Xi\frac{\tilde{\Omega}(\Xi)^2-3l^2\tilde{\Omega}'(\Xi)^2+2l^2\tilde{\Omega}(\Xi)\tilde{\Omega}''(\Xi)}{16G_{3}l\pi\tilde{\Omega}(\Xi)^2}\\
				&=\frac{\beta(3-l^2(16\pi^2\beta^2\gamma^2+3\mu^2+12\gamma(\pi\beta\mu-1)))}{24G_{3}l}.
			\end{align}
	The extra term compared with ADM gauge ($\mu=\gamma=0$) is:
		\begin{equation}
			-\frac{\beta l(16\pi^2\beta^2\gamma^2+3\mu^2+12\gamma(\pi\beta\mu-1))}{24G_{3}},
	\end{equation}
	that is to say, in generic gauges we can not define the holographic mass. Weyl factor means that we can introduce a source in the boundary CFT in the frame of holography.
	
 When the gravity is minimally coupled with other fields, we can draw on the examples of C-metric with hair in \cite{Arenas-Henriquez:2023hur} and consider the promotion of FG expansion in other situations. This problem will be discussed in the future.
	
	\bibliographystyle{unsrt}
	\bibliography{reference}
\end{document}